\documentclass[aps,prd,reprint, longbibliography,notitlepage,nofootinbib,nobibnotes,superscriptaddress,amsmath,amssymb,preprintnumbers]{revtex4-2}

\usepackage{graphicx} 
\usepackage{dcolumn}
\usepackage{bm}        
\usepackage{amssymb}   
\usepackage{adjustbox}  
\usepackage{amsmath,array}
\usepackage{comment}
\usepackage{natbib}
\usepackage{caption}
\usepackage{subcaption}
\usepackage[colorlinks=true,linkcolor=blue,citecolor=blue, urlcolor=blue]{hyperref}
\usepackage[capitalise]{cleveref}

\usepackage{graphicx}
\usepackage[usenames,dvipsnames]{color}
\usepackage[utf8]{inputenc}
\usepackage[T1]{fontenc}
\usepackage[english]{babel}
\usepackage{epigraph}
\usepackage{etoolbox}
\usepackage{ dsfont }
\usepackage{physics}
\usepackage{float}
\usepackage{xcolor}
\usepackage{siunitx}

\usepackage{xfrac}
\usepackage{amsmath,amssymb}
\usepackage{esdiff} 
\usepackage{commath} 
\usepackage{bbm} 
\usepackage{braket}
\usepackage{slashed}

\allowdisplaybreaks

\newcommand{\rmd}{\mathrm{d}}

\newcommand{\G}{\mathbf{G}}

\newcommand{\kompost}{K{\o}MP{\o}ST}

\newcommand{\xp}{\vec{x}}
\newcommand{\pp}{\vec{p}}

\newcommand{\xv}{\vec{x}}

\newcommand{\xT}{\mathbf{x}_T}			
\newcommand{\pT}{\mathbf{p}_T}			

\definecolor{oscar}{RGB}{22, 156, 172}
\definecolor{peter}{RGB}{255,0,0}

\definecolor{oscarC}{RGB}{22, 156, 172}

\makeatletter
\renewcommand\@makecaption[2]{%
  \par
  \vskip\abovecaptionskip
  \begingroup
   \small\rmfamily
    \begingroup
     \samepage
     \flushing
     \let\footnote\@footnotemark@gobble
     \@make@capt@title{#1}{#2}\par
    \endgroup
  \endgroup
  \vskip\belowcaptionskip
}
\makeatother

\begin{document}

\date{\today}

\title{Pre-equilibrium photons from the early stages of heavy-ion collisions}

\author{Oscar Garcia-Montero}
\email{garcia@physik.uni-bielefeld.de}
\affiliation{Fakult\"at f\"ur Physik, Universit\"at Bielefeld, D-33615 Bielefeld, Germany}

\author{Aleksas Mazeliauskas}
\email{a.mazeliauskas@thphys.uni-heidelberg.de}
\affiliation{Institut f\"ur Theoretische Physik, Universit\"at Heidelberg, 69120 Heidelberg, Germany}

\author{Philip Plaschke}
\email{pplaschke@physik.uni-bielefeld.de}
\affiliation{Fakult\"at f\"ur Physik, Universit\"at Bielefeld, D-33615 Bielefeld, Germany}

\author{S\"oren Schlichting}
\email{sschlichting@physik.uni-bielefeld.de}
\affiliation{Fakult\"at f\"ur Physik, Universit\"at Bielefeld, D-33615 Bielefeld, Germany}

\begin{abstract} 
We use QCD kinetic theory to compute photon production in the chemically equilibrating Quark-Gluon Plasma created in the early stages of high-energy heavy-ion collisions. We do a detailed comparison of pre-equilibrium photon rates to the thermal photon production. We show that the photon spectrum radiated from a hydrodynamic attractor evolution satisfies a simple scaling form in terms of the specific shear viscosity $\eta/s$ and entropy density $dS/d\zeta \sim {\scriptstyle \left(T\tau^{1/3}\right)^{3/2}}_\infty$. We confirm the analytical predictions with numerical kinetic theory simulations. We use the extracted scaling function to compute the pre-equilibrium photon contribution in $\sqrt{s_{NN}}=2.76\,\text{TeV}$ 0-20\% PbPb collisions. We demonstrate that our matching procedure allows for a smooth switching from pre-equilibrium kinetic to thermal hydrodynamic photon production. Finally, our publicly available implementation can be straightforwardly added to existing heavy ion models.
\end{abstract}

\maketitle

\tableofcontents

\section{Introduction}\label{sec:intro}
High-energy heavy-ion collisions produce an extremely hot and dense state of deconfined QCD matter. During the early stages of the collision, the QCD matter goes through a stage of kinetic and chemical equilibration, and the hydrodynamization of the Quark-Gluon Plasma (QGP) is swiftly achieved. Despite significant progress in the theoretical understanding of the early pre-equilibrium evolution~\cite{Schlichting:2019abc,Berges:2020fwq}, this stage is veiled from direct experimental observation using hadronic observables by the memory loss of the equilibrating medium and by the complicated nature of hadronization. However, electromagnetic probes, i.e., photons and dileptons, provide a unique tool to extract information about the pre-equilibrium epoch as they can escape the deconfined medium without rescattering~\cite{David:2019wpt,Monnai:2022hfs}.

Electromagnetic probes are produced during a heavy-ion collision through three main channels: hard scatterings in the first instants of the collision (prompt contribution), medium induced radiation (medium contribution), and the late-time hadronic decays (decay contribution). The sum of the first two channels (called the direct photons and dileptons) can be isolated by the subtraction of the decay products from the total yield. 

By now it has been well established, that in order to describe the experimentally measured direct photon spectra in heavy ion collisions, it is essential to include the in-medium radiation. Specifically, the medium-induced photon radiation dominates over the prompt photon production at low and intermediate transverse momenta~\cite{Turbide:2003si}. Beyond providing an additional source of photon production, the anisotropic expansion of the QGP correlates the photons to the collective flow of hadrons. This results in the well-known measurement of a non-zero photon elliptic flow~\cite{Chatterjee:2005de,Chatterjee:2013naa}, i.e., the second Fourier component in the azimuthal angle. So far, the simultaneous description of photon yield and elliptic flow by theoretical models has proved to be challenging, as photon radiation from the more isotropic and hotter earlier stages competes with the more anisotropic photon emission in the later colder stages of the collision. These apparent complex interplay is what has been dubbed the \textit{direct photon puzzle} in the literature~\cite{David:2019wpt, Monnai:2022hfs}. 

So far, the photon production during the pre-equilibrium stages has been neglected in most theoretical studies ~\cite{Shen:2013vja,Paquet:2015lta}, or described only in a simplified way~\cite{Monnai:2019vup,Monnai:2018eoh,Garcia-Montero:2019kjk,Garcia-Montero:2019vju}, where pre-equilibrium photon emission was computed using the parametric behavior in the `bottom-up` thermalization picture. Notable exceptions are provided by~\cite{Greif:2016jeb}, where the pre-equilibrium photon production was computed in the QCD transport approach BAMPS~\cite{Xu:2004mz}. Additionally, Ref.~\cite{Kasmaei:2019ofu} investigates the effects of momentum anisotropy on photon production. In Ref.~\cite{Gale:2021emg} the suppression of photon emission in gluon-dominated initial stages was estimated from the chemical equilibration rate in QCD kinetic theory simulations.



The central objective of this paper is to simultaneously achieve a consistent theoretical description of the kinetic and chemical equilibration of QGP and the photon production during the early pre-equilibrium stage based on leading order QCD kinetic theory~\cite{Arnold:2002ja,Arnold:2002zm,Arnold:2001ms,Arnold:2001ba}. Based on previous studies in strongly coupled holographic models of QGP and weakly coupled QCD kinetic theory, it has been observed that the early non-equilibrium evolution of QGP simplifies due to the phenomenon of hydrodynamic attractors (see Ref.~\cite{Florkowski:2017olj,Romatschke:2017ejr,Schlichting:2019abc,Berges:2020fwq,Soloviev:2021lhs} and references therein). In the transversely homogeneous and longitudinally boost-invariant Bjorken expansion, the equilibration of plasmas with different energy densities and coupling strengths can be universally described in terms of a single scaled time variable $\tilde{w} = \tau T/(4\pi \eta/s)$. Here the specific shear viscosity $\eta/s$ plays the role of coupling constant (stronger coupling---smaller viscosity). We will demonstrate that, under rather modest assumptions, the photon emission from the pre-equilibrium QGP also obeys a similar universal scaling relation. Specifically, the differential photon spectrum ${dN}/{d^2\xT d^2\pT dy}$ can be expressed in terms of a universal function $\mathcal{N}_\gamma$ of the scaled time variable $\tilde{w}$ and scaled transverse momentum $\bar{p}_T\propto {\sqrt{\eta/s} ~p_T}/{(s\tau)_\infty^{1/2}}$ with $(s\tau)_\infty$ being the entropy density per unit rapidity in the hydrodynamic phase, as
\begin{align}
    \frac{dN}{d^2\xT d^2\pT dy} = (\eta/s)^2 \Tilde{C}_\gamma^\text{ideal} \mathcal{N}_\gamma \left(\Tilde{w}, \bar{p}_T \right),\label{eq:photonscaling}
\end{align}
where $\Tilde{C}_\gamma^\text{ideal}$ is a numerical constant describing the thermal photon production rate. 
\Cref{eq:photonscaling} along with the calculation of the scaling function $\mathcal{N}_\gamma$ in leading order QCD kinetic theory are the main results of our paper\footnote{ Even though in this work \cref{eq:photonscaling} is obtained self-consistently within leading order QCD kinetic theory, we believe that it is also sensible to employ this expression for estimates of the pre-equilibrium photon production in different microscopic theories, e.g., by supplying the next-to-leading order photon rate~\cite{Ghiglieri:2013gia} in the calculation of the coefficient $\Tilde{C}_\gamma^\text{ideal}$ to (partially) account for higher order corrections.}.

Based on \cref{eq:photonscaling} it becomes straightforward to compute the
photon emission during the pre-equilibrium QGP stage, which smoothly matches to the thermal photon production in hydrodynamic simulations of equilibrated QGP. By performing this calculation explicitly, we find that although the pre-equilibrium photons make up only a small fraction of the total photon yield, their contribution at intermediate momenta $p_T\sim 3\,\text{GeV}$ can be dominant, such that a careful study of this momentum region may have the potential to differentiate between different QGP thermalization scenarios.

 This paper is organized as follows: in \cref{sec:pre-eqPhoton} we present the relevant aspects of the pre-equilibrium photon production in the Effective Kinetic Theory framework. In \cref{sec:ScalingOfYield} we derive a universal scaling function of the time-integrated photon spectrum. 
This formula is the most important result of this paper for phenomenological computations of photon production.
 \Cref{sec:results} is dedicated to the detailed comparison of non-equilibrium and thermal photon production. In particular, we verify the predicted scaling of the time-integrated photon spectrum.
 Using the scaling formula in \cref{sec:pheno} we add the pre-equilibrium photons to other sources of photon production in heavy-ion collisions and compare to the experimental data. We also study the sensitivity of the in-medium photon yield to the hydrodynamization time, and find it to be robust to the variation of this parameter. We conclude with a summary and outlook in \cref{sec:conclusion}. In \cref{sec:TheoryNonEqQGP} we document the collision process of quarks and gluons, while in \cref{sec:sourcesPheno} we detail the prompt and thermal photon computations for phenomenological comparisons.

We note that, throughout this manuscript, we denote four vectors as $P$ or $p^\mu$, three vectors as $\pp$, (transverse) two vectors as $\pT$, their magnitude as $p_T$ and energy as $E_p=p^0=|\pp|=p$.

\section{Photon production in QCD kinetic theory}\label{sec:pre-eqPhoton}
We employ the high temperature effective kinetic theory of non-Abelian gauge theories (AMY EKT)~\cite{Arnold:2002zm} to describe the pre-equilibrium evolution of the QGP. The dynamics of color, flavor, spin, and polarization averaged single particle distribution function $f_a(X,\pp)$ is governed by the Boltzmann equation
\begin{align}
  \left[ \frac{\partial}{\partial t} + \frac{\pp}{p} \cdot \nabla_{\xv} \right] f_a(X,\pp) =  C^{2 \leftrightarrow 2}_a [f_a](X,\pp) + C^{1 \leftrightarrow 2}_a [f_a](X,\pp) \, ,\label{eq:boltz}
\end{align}
where the collision integral $C_a^{2 \leftrightarrow 2}$ describes the elastic $2 \leftrightarrow 2$ processes, $C_a^{1 \leftrightarrow 2}$ encodes contributions from effective inelastic $1 \leftrightarrow 2$ processes\footnote{Our collision integrals come with a positive sign in \cref{eq:boltz} and therefore differ from Ref.~\cite{Arnold:2002zm} by the overall minus sign.}. The subscript $a = g, q, \gamma$ indicates the particles species. In this work, $f_q$ represents a fermion distribution averaged over $N_f=3$ flavors, $N_c=3$ colors, and two spin states of massless quarks and anti-quarks (see \cref{tab:dof}).
\begin{table}[t]
    \centering
    \begin{tabular}{c|c|c|c}
        $s$ & $d_s$ & $\nu_s$ &  $C_s$ \\
        \hline
        $g$ &  8 & 16 &  3\\
        $q$ &  3 & 6 &  $\frac{4}{3}$\\
        $\gamma$ &  1 & 2 &  0
    \end{tabular}
    \caption{Species $(s)$, dimension of the color representation ($d_s$), number of degrees of freedom $(\nu_s)$ and Casimirs $(C_s)$ for gluons, quarks and photons.}
    \label{tab:dof}
\end{table}

In the case of photons, we can relate the collision integrals to the local production rate $dN_\gamma/d^4Xd^3\vec{p}$ according to
\begin{align}
   \frac{dN_\gamma}{d^4X d^3\pp} = \frac{\nu_\gamma}{(2\pi)^3} C_{\gamma}^{1\leftrightarrow 2}[f](x,\pp) + \frac{\nu_\gamma}{(2\pi)^3} C_{\gamma}^{2\leftrightarrow 2}[f](x,\pp) \, ,
\end{align}
while the distribution function $f_\gamma$ is given as
\begin{align}
  \frac{dN_\gamma}{d^3\xp d^3\pp} = \frac{\nu_\gamma}{(2\pi)^3}  f_{\gamma}(X,\pp) \, .
\end{align}
We will follow previous works in QCD kinetic theory~\cite{Kurkela:2018oqw,Du:2020dvp}, where transverse inhomogeneities and transverse expansion were neglected during the early pre-equilibrium phase. We thus focus on homogeneous, boost-invariant systems for which the phase-space density only depends on local momenta $P = (p_T\cosh(y-\zeta), \pT, p_T\sinh(y-\zeta))$ and Bjorken time $\tau=\sqrt{t^2-z^2}$. Here $y=\text{artanh}(p^z/p)$ is the momentum-space rapidity and $\zeta=\text{artanh}(z/t)$ the space-time rapidity. Since momentum and space-time rapidities only appear as a difference, we can trade the $\zeta$ integral for an integral over the longitudinal momentum $p^z=p_T \sinh(y-\zeta)$ in the local rest frame and the Boltzmann equation simplifies to
\begin{align}\label{eq:FinalBoltzmannEquation}
     \left[ \frac{\partial}{\partial \tau} - \frac{p^z}{\tau} \frac{\partial}{\partial p^z} \right] f_a(\tau,\pp) =  C^{2 \leftrightarrow 2}_a [f_a](\tau,\pp) + C^{1 \leftrightarrow 2}_a [f_a](\tau,\pp) \, .
\end{align}

Our kinetic description includes all leading-order processes for quarks, gluons, and photons. The in-medium scattering matrix elements for the elastic collisions are evaluated using isotropic screening regulator~\cite{Kurkela:2015qoa,Kurkela:2018vqr,Kurkela:2014tea,Kurkela:2018oqw,Du:2020dvp}. For the inelastic collisions, the in-medium splitting rates are found by solving an integral equation reproducing the Landau-Pomeranchuk-Migdal (LPM) effect. 
In the next section, we specify the relevant collision integrals for photon production: $2 \leftrightarrow 2$ processes (Compton scattering, elastic pair annihilation) and effective collinear $1 \leftrightarrow 2$ processes (Bremsstrahlung, inelastic pair annihilation). The analogous collision integrals for QCD processes have been documented in the previous works~\cite{Du:2020dvp} and are given in \cref{sec:TheoryNonEqQGP} for completeness.

The kinetic processes would reach thermal equilibrium in an infinite system, including photon degrees of freedom. In realistic heavy ion collisions, the QGP fireball size $\sim 10-20\,\text{fm}$ is much smaller than $\sim 300-500\,\text{fm}$ needed for photons to re-interact~\cite{Paquet:2016ulk}.
Therefore in photon collision integrals, we will not keep the detailed balance and will only consider the gain terms. For the same reason, we will neglect the effect of Bose enhancement on photons.


\subsection{Elastic processes}\label{sec:elastic}
\begin{figure}[h!]
    \begin{minipage}{0.2\textwidth}
    \centering
    \includegraphics[width=0.9\textwidth]{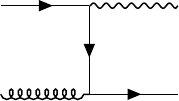}
    \end{minipage}
    \begin{minipage}{0.2\textwidth}
    \centering
    \includegraphics[width=0.9\textwidth]{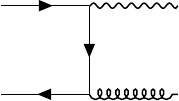}
    \end{minipage}
    \caption{The $t$-channels of elastic processes contributing to photon production at leading order. \textsl{Left}: Compton scattering. \textsl{Right}: Elastic pair annihilation. The $s$- and $u$-channels are not shown but included in the calculations.}
    \label{fig:ElasticProcessesDiagrams}
\end{figure}

The elastic processes included in our calculation are elastic pair annihilation (EPA) of quarks and antiquarks and Compton scattering (CS), where a quark/antiquark scatters with a gluon, see \cref{fig:ElasticProcessesDiagrams}. 

The corresponding collision integrals are given in AMY formalism \cite{Arnold:2002zm} as
\begin{subequations}
    \begin{align}
        &C^{2 \leftrightarrow 2}_{\gamma,\text{EPA}} [f](\pp_3) = \frac{1}{2\nu_\gamma} ~\frac{1}{2{p_3}} \\
        &\int d\Pi_{2 \leftrightarrow 2} \left| \mathcal{M}^{q\Bar{q}}_{\gamma g}(\pp_1,\pp_2|\pp_3,\pp_4) \right|^2 f_q(\pp_1) f_q(\pp_2) (1 + f_g(\pp_4)) \, , \nonumber \\
        &C^{2 \leftrightarrow 2}_{\gamma,\text{CS}} [f](\pp_3) = \frac{1}{2\nu_\gamma} ~\frac{1}{2{p_3}} \\
        &\int d\Pi_{2 \leftrightarrow 2} \left| \mathcal{M}^{gq}_{\gamma q}(\pp_1,\pp_2|\pp_3,\pp_4) \right|^2 f_g(\pp_1) f_q(\pp_2) (1 - f_q(\pp_4)) \, . \nonumber
\end{align}
\end{subequations}
The matrix elements are summed over the degrees of freedom of incoming and outgoing particles and are given in Mandelstam variables by~\cite{Ghiglieri:2020dpq}
\begin{subequations}
    \begin{align}
        \left| \mathcal{M}^{q\Bar{q}}_{\gamma g}(\pp_1,\pp_2|\pp_3,\pp_4) \right|^2 &= 8e^2 \sum_s q_s^2 g^2 d_F C_F \left[ \frac{-s}{\underline{t}} + \frac{t}{-s} \right] \,\label{eq:MEPA} ,\\
        \left| \mathcal{M}^{gq}_{\gamma q}(\pp_1,\pp_2|\pp_3,\pp_4) \right|^2 &= 8e^2 \sum_s q_s^2 g^2 d_F C_F \left[ \frac{u}{\underline{t}} + \frac{t}{\underline{u}} \right] \, ,\label{eq:MCS}
\end{align}
\end{subequations}
where the quark degrees of freedom are $d_F=N_c=3$, fundamental Casimir $C_F=(N_c^2-1)/(2N_c)=4/3$, electromagnetic coupling constant $e^2\approx 4\pi/137$, strong coupling constant $g^2 = 4\pi \alpha_s = \lambda/N_c$, and the sum of electric charges for $N_f=3$ flavours of quarks is
$\sum_{s} q_s^2 = \frac{2}{3}$.

Because of the enhancement of soft $t$- and $u$-channel exchanges, the vacuum matrix elements \cref{eq:MEPA,eq:MCS} would be divergent. Inside the medium, this is regulated by screenings effects through the insertions of Hard-Thermal loop (HTL) self-energies~\cite{Arnold:2002zm}. Implementation of full HTL results is complicated~\cite{Kasmaei:2019ofu,Schenke:2006yp}%
, so instead, we use isotropic screening~\cite{Kurkela:2015qoa,Kurkela:2018vqr,Kurkela:2014tea,Kurkela:2018oqw}, where the soft $t$- and $u$-channels are regulated by the screening mass $m^2_{q} =g^2 C_F \int \frac{d^3\vec{p}}{(2\pi)^3} \frac{1}{2p} \left[ 2 f_g(\pp) + 2f_{q}(\pp) \right]$. Namely, we substitute 
\begin{align}
    \underline{t} = t \left( 1 + \frac{\xi_q^2m_q^2}{\vec{q}_t^2} \right) \qquad , \qquad \underline{u} = u \left( 1 + \frac{\xi_q^2m_q^2}{\vec{q}_u^2} \right)
\end{align}
with $\vec{q}_{u/t}$ being the momentum of the exchange particle and $\xi_q=e/\sqrt{2}$ was determined in \cite{Ghiglieri:2015ala, Kurkela:2018oqw} to reproduce gluon to quark conversion $gg\to q\bar{q}$ at leading order for isotropic distributions.

\subsection{Inelastic processes}\label{sec:PhotonProductionFromInelasticProcesses}
\begin{figure}[h!]
    \begin{minipage}{0.2\textwidth}
    \centering
    \includegraphics[width=0.9\textwidth]{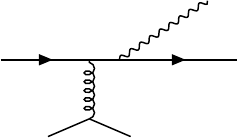}
    \end{minipage}
    \begin{minipage}{0.2\textwidth}
    \centering
    \includegraphics[width=0.9\textwidth]{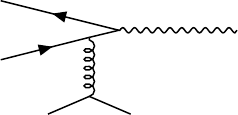}
    \end{minipage}
    \caption{Inelastic processes contributing to photon production at leading order. \textsl{Left}: Bremsstrahlung. \textsl{Right}: Inelastic pair annihilation.}
    \label{fig:InelasticProcessesDiagrams}
\end{figure}

A full leading-order description of photon production includes inelastic contributions coming from inelastic pair annihilation (IPA) and Bremsstrahlung (BS) off quarks and antiquarks. The corresponding Feynman diagrams are presented in \cref{fig:InelasticProcessesDiagrams}.

The collision integrals are obtained from the AMY formalism \cite{Arnold:2002zm}. At leading order inelastic processes can be taken to be collinear. Therefore we follow \cite{Du:2020dvp} and rewrite the integrals as an effectively one-dimensional integral in terms of the collinear splitting ($z$) and joining ($\Bar{z} = 1-z$) fractions
\begin{subequations}
    \begin{align}
        C_{\gamma, \text{IPA}}^{1 \leftrightarrow 2}[f](p) &=  \frac{2\sum\nolimits_{s} q_s^2}{2\nu_\gamma} \int_{0}^{1} \dd{z} \nu_\gamma \frac{\dd{\Gamma_{q\overline{q}}^{\gamma}}}{\dd{z}}(p,z) f_q(zp) f_q(\overline{z}p) \, , \nonumber \\
        C_{\gamma, \text{BS}}^{1 \leftrightarrow 2}[f](p) &=  \frac{2\sum\nolimits_{s} q_s^2}{2\nu_\gamma} \\
       \int_{0}^{1}& \dd{z} \nu_q \bigg[ \frac{1}{z^3} \frac{\dd{\Gamma_{\gamma q}^{q}}}{\dd{z}}\left(\textstyle \frac{p}{z},z \right) f_q \left(\textstyle \frac{p}{z} \right) \left( 1-f_q \left(\textstyle \frac{\overline{z}}{z}p \right) \right) \nonumber \\
        + &\frac{1}{\overline{z}^3} \frac{\dd{\Gamma_{\gamma q}^{q}}}{\dd{z}}\left(\textstyle \frac{p}{\overline{z}},\overline{z} \right) f_q \left(\textstyle \frac{p}{\overline{z}} \right) \left( 1-f_q \left(\textstyle \frac{z}{\overline{z}}p \right) \right) \bigg] \, .
    \end{align}
\end{subequations}
The effective inelastic rate $\frac{d\Gamma_{bc}^{a}}{dz}$ is obtained by considering the overall probability of a single radiative emission/absorption over the course of multiple successive elastic interactions. The destructive interference of these scatterings leads to the suppression of radiative emission/absorption, known
as the LPM effect. The splitting probability is reduced to an effective collinear rate by integrating over the parametrically small transverse momentum accumulated during the emission/absorption process. It can be expressed accordingly
\begin{align}
    \frac{d\Gamma_{bc}^{a}}{dz} (p,z) = \frac{\alpha_{\text{EM}} P_{bc}^{a}(z)}{[2pz(1-z)]^2} \int \frac{d^2\mathbf{p}_b}{(2\pi)^2} \Re{2\mathbf{p}_b \cdot \G(\mathbf{p}_b)} \, ,\label{eq:rateGamma}
\end{align}
where $\alpha_s = \frac{\lambda}{4 \pi N_c} = \frac{g^2}{4 \pi}$ and $P_{bc}^{a}$ is the leading-order QCD splitting function~\cite{Ellis:1996mzs}
\begin{subequations}
    \begin{align}
        P^\gamma_{q\Bar{q}} (z) &= N_c ~( (1-z)^2 + z^2 ) \, , \\
        P^q_{\gamma q} (z) &= \frac{1 + (1-z)^2}{z} = P^q_{q \gamma} (1-z) \, .
\end{align}
\end{subequations}
The term $\Re{2\mathbf{p}_b \cdot \G(\mathbf{p}_b)}$ in \cref{eq:rateGamma} captures the relevant aspects of the current-current correlation function inside the medium to treat the LPM-effect via effective vertex resummation. It satisfies the integral equation
\begin{align}\label{eq:IntegralEquationEffectiveRate}
    2\mathbf{p}_b& = i\delta E(\mathbf{p}_b) \G(\mathbf{p}_b)+ \frac{1}{2} \int \frac{d^2\mathbf{q}}{(2\pi)^2} ~\frac{d\Bar{\Gamma}^{\text{el}}}{d^2\mathbf{q}} \times \nonumber \\
    &\Big\{ \left( C^R_b + C^R_c - C^R_a \right) \left[ \G(\mathbf{p}_b) - \G(\mathbf{p}_b-\mathbf{q}) \right] \nonumber \\
    &\left( C^R_c + C^R_a - C^R_b \right) \left[ \G(\mathbf{p}_b) - \G(\mathbf{p}_b-z\mathbf{q}) \right] \\
    &\left( C^R_a + C^R_b - C^R_c \right) \left[ \G(\mathbf{p}_b) - \G(\mathbf{p}_b-(1-z)\mathbf{q}) \right] \Big\} \nonumber \, ,
\end{align}
where
\begin{align}
    \delta E(\mathbf{p}_b) = \frac{\mathbf{p}_b^2}{2pz(1-z)} + \frac{m^2_{\infty,b}}{2zp} + \frac{m^2_{\infty,c}}{2(1-z)p} - \frac{m^2_{\infty,a}}{2p} \, .
\end{align}
Here, $m^2_{\infty,a}$, $m^2_{\infty,b}$, $m^2_{\infty,c}$ denotes the asymptotic masses of the different particles, namely, $m^2_{\infty,{q}} = 2 m_{q}^2$ for quarks and $m^2_{\infty,\gamma} = 0$ for photons. 
The thermal quark mass $m_{q}$ is given by
\begin{align}
    m_{q}^2 &= g^2 C_F \int \frac{d^3\vec{p}}{(2\pi)^3} \frac{1}{2p} \left[ 2 f_g(\pp) + 2f_{q}(\pp) \right] \, ,
\end{align}
where we used $f_{q} = f_{\bar{q}}$ for a system without net charges.

The Casimir of the representation $C^R_a$, $C^R_b$, $C^R_c$ are given by $C^R_g = C_A$ for gluons, $C^R_q = C_F$ for quarks and $C_\gamma = 0$ for photons. Finally $d\Bar{\Gamma}^{\text{el}}/d^2\mathbf{q}$ is the differential elastic scattering rate, which is stripped of its color factor and at leading order is given by~\cite{Du:2020dvp}
\begin{align}
    \frac{d\Bar{\Gamma}^{\text{el}}}{d^2\mathbf{q}} = g^2 T^\ast \frac{m_D^2}{\mathbf{q}^2 ( \mathbf{q}^2 + m_D^2 )} \, 
\end{align}
with the Debye mass
\begin{align}
    m_D^2 &= 4g^2 \int \frac{d^3\vec{p}}{(2\pi)^3} \frac{1}{p} \left[ N_cf_g(\pp) + N_f f_{q}(\pp) \right] \, .
\end{align}
The effective inelastic in-medium rates are sensitive to the density of elastic interaction partners
\begin{align}
    m_D^2 T^\ast = 2g^2\int \frac{d^3\vec{p}}{(2\pi)^3} \bigg[ &N_c f_g(\pp) (1 + f_g(\pp)) \nonumber \\
    +  &N_f f_{q}(\pp) (1 - f_{q}(\pp)) \bigg] \, ,
\end{align}
due to the fact that inelastic interactions are induced by elastic interactions. The quantity $g^2T^\ast$ in \cref{eq:IntegralEquationEffectiveRate} characterizes then the rate of small angle scatterings in the plasma where, $T\ast$ is defined such that in equilibrium it corresponds to the equilibrium temperature $T^\ast \overset{\text{eq}}{=} T_\text{eq}$.

Solutions to \cref{eq:IntegralEquationEffectiveRate} are found by first performing a Fourier transform to the impact parameter space. One solves the resulting ordinary differential equation with specified boundary conditions~\cite{Anisimov:2010gy,Moore:2021jwe}. For details of numerical implementation, we refer to the previous work~\cite{Du:2020dvp,Moore:2021jwe}.

\begin{figure*}
    \centering
    \includegraphics[width=0.45\textwidth]{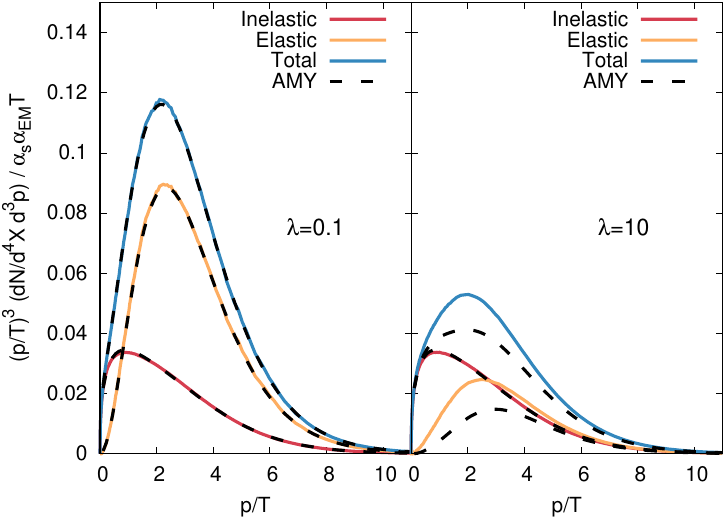}
    \includegraphics[width=0.45\textwidth]{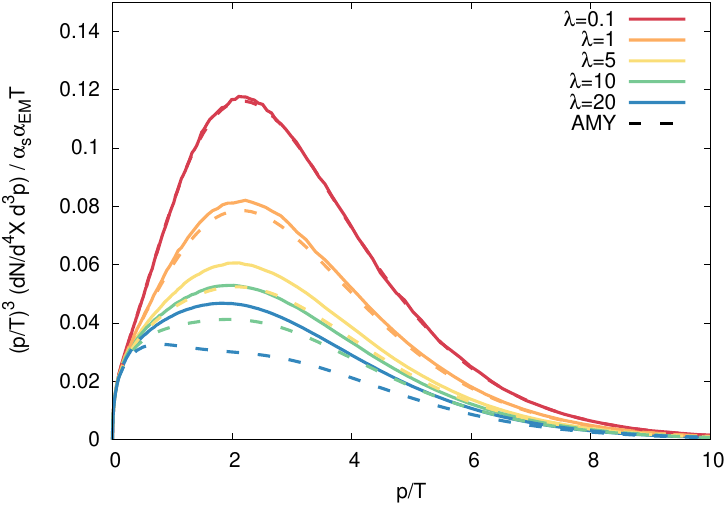}
    \caption{Comparison of photon production rates in EKT (solid lines) with semi-analytical AMY results (dashed lines)~\cite{Arnold:2001ms}. (Left) Inelastic, elastic and total photon rates for $\lambda=0.1$ and $\lambda=10$, (right) total photon rate for $\lambda=0.1,1,5,10,20$.}
    \label{fig:AMYrates}
\end{figure*}
\subsection{Comparison to equilibrium AMY rates}

Before we proceed to the calculation of the pre-equilibrium photon production, we compare our numerical implementation of photon rates of \cref{sec:elastic,sec:PhotonProductionFromInelasticProcesses} to the equilibrium parametrizations of photon rates given in Ref.~\cite{Arnold:2001ms}. In \cref{fig:AMYrates} we present the results for different coupling strengths $\lambda=g^2N_c$. In the left panel of \cref{fig:AMYrates} we compare different photon production channels. For small coupling, $\lambda=0.1$, we find a perfect agreement between our and AMY results for all channels. We also note that the elastic processes are the dominant photon production channel for this coupling except for the low momentum region $p<T$. For larger coupling, $\lambda=10$, the relative contribution of the elastic channel is greatly suppressed and becomes equal to the inelastic channel even for $p>T$. In this case, we see the difference in elastic processes between our and AMY calculations. We note that authors of~\cite{Arnold:2001ms} use different infrared regularisation in the elastic channel, so small differences are to be expected. The difference becomes more pronounced for larger values of the coupling where the subleading corrections to the infrared regulators become apparent. A similar trend is seen in the right panel of \cref{fig:AMYrates} where we show just the total photon rate for different couplings $\lambda=0.1,1,5,10,20$.

\section{Scaling laws for photon spectrum}\label{sec:ScalingOfYield}

The thermalization of QGP in QCD kinetic theory has been studied extensively~\cite{Kurkela:2015qoa,Keegan:2015avk,Kurkela:2018wud,Kurkela:2018vqr,Kurkela:2018oqw,Kurkela:2018xxd,Mazeliauskas:2018yef,Almaalol:2020rnu,Du:2020dvp,Du:2020zqg,Du:2022bel}. One of the key discoveries is the emergence of so-called hydrodynamic attractors. Namely, the key bulk characteristics of the QGP, like energy or pressure anisotropy, approximately follow a universal function, which is a function only of the scaled time $\tilde{w}=\tau T/(4\pi\eta/s)$. For example, the energy density $e(\tau)$ in Bjorken-expansion follows
\begin{align}
    \tau^{4/3} e(\tau) = \mathcal{E}(\Tilde{w}) \left(\tau^{4/3} e \right)_\infty,\label{eq:attractor}
\end{align}
where $\left(\tau^{4/3} e \right)_\infty$ is the asymptotic equilibrium constant and $\mathcal{E}(\tilde{w})$ is a universal function of scaled time $\tilde w$~\cite{Giacalone:2019ldn}.
This is the expected hydrodynamic behavior near equilibrium. The collapse of simulations with different initial conditions to the same curve $\mathcal{E}$ in deeply anisotropic regime $|(T^{zz}-T^{xx})/T^{00}|\sim \mathcal{O}(1)$ is called the hydrodynamic (energy) attractor.

The simplicity of non-equilibrium system following the attractor allowed analytical computation of entropy production in the non-equilibrium stages~\cite{Giacalone:2019ldn}. Scaling with $\tilde w$ has also been observed in the chemical equilibration of QGP~\cite{Kurkela:2018oqw,Kurkela:2018xxd}, and recently been used to estimate the dilepton production during the pre-equilibrium stage~\cite{Coquet:2021gms,Coquet:2021lca}. It is thus reasonable to expect that photon radiation off equilibrating fermions might also follow $\tilde w$-scaling. In this section, we will derive the analytical formulas for the photon spectra from expanding QGP.

\subsection{Scaling for evolution along a hydrodynamic attractor}\label{sec:NonEqScalingOfYield}

The Lorentz invariant photon spectrum radiated by the QGP is given by the volume integral
\begin{align}
    E_{p} \frac{dN_\gamma}{d^3\pp} = \frac{\nu_\gamma}{(2\pi)^3} \int d^4X  p^{\mu}\partial_{\mu} f_{\gamma}(X,\pp),\label{eq:spectra1}
\end{align}
where the phase-space density of photons is defined as
\begin{align}
  \frac{\nu_\gamma}{(2\pi)^3}  f_{\gamma}(X,\pp) \equiv \frac{dN_\gamma}{d^3\xp d^3\pp} \,
\end{align}
and as before, the collision integral is determined from
\begin{align}
    \frac{dN_\gamma}{d^4X d^3\pp} = \frac{\nu_\gamma}{(2\pi)^3} C_{\gamma}[f](X,\pp) \, .
\end{align}
The integral \cref{eq:spectra1} is similar to the Cooper-Frye integral for hadrons, but the surface integral is replaced by a volume integral.

As the momentum differential can be rewritten according to $d^3\vec{p} = d^2\pT dy$, where $y$ is the momentum space-rapidity, the transverse momentum spectrum is derived by inserting the Boltzmann equation 
\begin{align}
    p^{\mu}\partial_{\mu} f_{\gamma}(X,\pp) = E_p C_{\gamma}[f](X,\pp)
\end{align}
into \cref{eq:spectra1} and is determined according to
\begin{align}
     \frac{dN_\gamma}{d^2\mathbf{p}_T dy}  =\frac{\nu_\gamma}{(2\pi)^3} \int d^4X  p\, C_{\gamma}[f](X,\pp) \, ,
\end{align}
where we combined elastic and inelastic processes in a single integral $C_\gamma$. Similar to the discussion around \cref{eq:FinalBoltzmannEquation}, for a boost-invariant expansion and transversely homogeneous expansion, the photon emission rate $C_{\gamma}$ is a function of the Bjorken time $\tau$ and local momentum
$P = (p_T\cosh(y-\zeta), \pT, p_T\sinh(y-\zeta))$ and we can express the $\zeta$ integral for an integral over the longitudinal momentum $p^z=p_T \sinh(y-\zeta)$ in the local rest frame, such that
\begin{align}
     \frac{dN_\gamma}{d^2\xT d^2\pT dy}  =\frac{\nu_\gamma}{(2\pi)^3} \int_{\tau_\text{min}}^{\tau_\text{max}}\!\!\! d\tau \tau \int d{p}^z C_{\gamma}[f](\tau,\pp) ,\label{eq:spectra2}
\end{align}
Now if we assume that the phase-space distributions of quarks and gluons obey $\tilde w$-scaling, such that in the local rest frame $f_{a}(X,P)=f_{a}(\tilde{w},\vec{p}/T_{\rm eff}(\tau))$, we can write the collision integral as a dimensionless function $\Bar{C}_\gamma$ of $\tilde w$ and scaled momenta
\begin{align}
    C_\gamma(\tau, \pp) = T_{\text{eff}}(\tau) \Bar{C}_\gamma \left(\Tilde{w}, \frac{\pp}{T_\text{eff}(\tau)} \right) \, ,\label{eq:scalingofC}
\end{align}
where the effective temperature is defined by the Landau-matching condition
\begin{align}
    e(\tau) \equiv \frac{\pi^2}{30} \nu_\text{eff} T_\text{eff}^4(\tau) \, ,\label{eq:landau}
\end{align}
with $\nu_\text{eff} = \nu_g + 2 N_f \frac{7}{8} \nu_q$. Using \cref{eq:attractor,eq:landau} and the definition of $\tilde w$, we can write the effective temperature and Bjorken time in terms of $\tilde w$ variable
\begin{subequations}
\begin{align}
   & T_\text{eff}(\tau) = \frac{\left(\tau^{1/3} T \right)^{3/2}_\infty}{(4\pi\eta/s)^{1/2}} \frac{\mathcal{E}^{3/8}(\Tilde{w})}{{\tilde w}^{1/2}}\, \label{eq:Ttow}\\
&\tau = \frac{(4\pi \eta/s) \tilde w}{T_\text{eff}}=\frac{(4\pi \eta/s)^{3/2}}{\left(\tau^{1/3} T \right)^{3/2}_\infty} \frac{{\tilde w}^{3/2} }{\mathcal{E}^{3/8}(\Tilde{w})},\label{eq:tautow}
\end{align}
\end{subequations}
where 
\begin{subequations}
\begin{align}
    (\tau^{4/3} e )_\infty&\equiv \frac{\pi^2}{30} \nu_\text{eff} (\tau^{1/3} T )^4_\infty,\\
    (\tau s )_\infty&\equiv \frac{4\pi^2}{90} \nu_\text{eff} (\tau^{1/3} T )^3_\infty.
\end{align}
\end{subequations}

By plugging \cref{eq:scalingofC,eq:Ttow,eq:tautow} into \cref{eq:spectra2} we then obtain the scaling formula for photon spectrum emitted by the equilibrating QGP
\begin{align}
  &\frac{dN_\gamma}{d^2\xT d^2\pT dy} \nonumber \\
 &=\frac{\nu_\gamma(4\pi \eta/s)^2 }{2(2\pi)^3}\int_{\scriptscriptstyle \frac{\tilde w^3_\text{min}}{\mathcal{E}^{3/4}(\tilde w_\text{min})}}^{\scriptscriptstyle \frac{\tilde w^3_\text{max}}{\mathcal{E}^{3/4}(\tilde w_\text{max})}} d\left(\frac{\tilde w^3}{\mathcal{E}^{3/4}(\Tilde{w})}\right) \frac{\mathcal{E}^{3/8}(\tilde{w})}{{\tilde w}^{1/2}}\nonumber \\ &\sqrt{4\pi} \int d\bar{p}^z 
     \bar C_{\gamma}[f]\left(\tilde w,\frac{\sqrt{4\pi\tilde w}}{\mathcal{E}^{3/8}(\tilde{w})} \bar{\mathbf{p}}_T,\frac{\sqrt{4\pi\tilde w}}{\mathcal{E}^{3/8}(\Tilde{w})} \bar{p}^z\right) ,\label{eq:photonattractor}
\end{align}
where we defined $\bar{\mathbf{p}}_T ={(\eta/s)^{1/2}} \pT/{\scriptstyle \left(\tau^{1/3} T \right)^{3/2}_\infty}$ and similarly $\bar{p}^z={(\eta/s)^{1/2}} p^z/{\scriptstyle \left(\tau^{1/3} T \right)^{3/2}_\infty}$. 

We note that the photon production rate $\Bar{C}_\gamma$ itself also explicitly contains the strong coupling constant (see \cref{sec:elastic,sec:PhotonProductionFromInelasticProcesses}). However, by looking at the ratio of the non-equilibrium rate to the thermal rate, this additional coupling dependence can be cancelled out.
By following this logic and taking the limit $\tilde{w}_\text{min} \rightarrow 0$ we define\footnote{Although the QCD kinetic simulation starts at a finite time, we can take the limit $\tilde{w}_\text{min} \rightarrow 0$ here, as initially there are no quarks in the system and as such there are no photons produced at these times.}
\begin{align}
\mathcal{N}_{\gamma}(\tilde{w}_\text{max},\bar{\mathbf{p}}_T) &=    \sqrt{\frac{4}{\pi}}\frac{\nu_\gamma}{\tilde{C}_\gamma^\text{ideal}}\int_{0}^{\scriptscriptstyle \frac{\tilde w^3_\text{max}}{\mathcal{E}^{3/4}(\tilde w_\text{max})}} d\left(\frac{\tilde w^3}{\mathcal{E}^{3/4}(\Tilde{w})}\right) \frac{\mathcal{E}^{3/8}(\Tilde{w})}{{\tilde w}^{1/2}}\nonumber \\ 
\int &d\bar{p}^z 
     \bar C_{\gamma}[f]\left(\tilde w,\frac{\sqrt{4\pi\tilde w}}{\mathcal{E}^{3/8}(\Tilde{w})} \bar{\mathbf{p}}_T,\frac{\sqrt{4\pi\tilde w}}{\mathcal{E}^{3/8}(\Tilde{w})} \bar{p}^z\right)
\end{align}
where
\begin{align}
\label{eq:CIdealDef}
    \tilde{C}_\gamma^\text{ideal} = 4 \int_0^\infty d\left(\frac{p}{T}\right) ~\left(\frac{p}{T}\right)^4 \Bar{C}^{\rm eq}_\gamma \left(\frac{p}{T}\right) \, ,
\end{align}
is obtained as a moment of the equilibrium photon rate (see next section).
We now see that as advertised in \cref{sec:intro} the photon spectrum
\begin{align}
     \frac{dN_\gamma}{d^2\xT d^2\pT dy} = (\eta/s)^2~\tilde{C}_\gamma^\text{ideal}~\mathcal{N}_{\gamma}(\tilde{w},\bar{\mathbf{p}}_T)
\end{align}
is naturally written in terms of the scaled transverse momentum $\bar{\mathbf{p}}_T = (4\nu_\text{eff}\pi^2/90)^{1/2}(\eta/s)^{1/2} \pT/(s T)_\infty^{1/2} $ and exhibits an overall $(\eta/s)^2$ scaling.

\subsection{Scaling for ideal Bjorken expansion}\label{sec:Bjorken}

Starting from \cref{eq:spectra2} we can also derive the idealized thermal spectrum if we assume the system is in equilibrium for all times. In this case $T_\text{eff}=T(\tau) =(\tau^{1/3}T)_\infty/\tau^{1/3} $ and the collision integral only depends on $z=|\pp|/T(\tau) = p_T\cosh \zeta/T(\tau)$, such that
\begin{align}
    & \frac{dN_\gamma}{d^2\xT d^2\pT dy}  =\frac{\nu_\gamma}{(2\pi)^3} \frac{3(\tau^{1/3}T)^6_\infty}{p_T^4}  \int_{-\infty}^\infty \frac{d\zeta}{\cosh^4 \zeta}\nonumber \\
     &\quad\int_{0}^{\infty}\!dz\Theta(z-z_\text{min})\Theta(z_\text{max}-z)   z^4   \bar{C}_{\gamma}[f](z),
\end{align}
where the step function $\Theta$ depends on the starting and final temperatures of Bjorken expansion.
\begin{equation}
z_\text{min} = \frac{p_T\cosh \zeta}{T(\tau_\text{min})}, \quad z_\text{max} = \frac{p_T\cosh \zeta}{T(\tau_\text{max})}.
\end{equation}
By interchanging the order of integration, we can find the integration limits on the rapidity integral. For convenience, we introduce a change of variables $u=\cosh \zeta$. We split the $z$ integral into three cases: $z<p_T/T(\tau_\text{min})$, $p_T/T(\tau_\text{min})<z<p_T/T(\tau_\text{max})$ and $z>p_T/T(\tau_\text{max})$. Because $u>1$, the first case does not satisfy $z_\text{min}<z<z_\text{max}$. The other two cases give
\begin{align}
     &\frac{dN_\gamma}{d^2\xT d^2\pT dy}  =\frac{\nu_\gamma}{(2\pi)^3} \frac{3(\tau^{1/3}T)^6_\infty}{p_T^4}2 \Big [ \nonumber \\
     &\int_{p_T/T(\tau_\text{max})}^\infty dz  z^4  \bar{C}_{\gamma}[f](z) \int_{z T(\tau_\text{max})/p_T}^{z T(\tau_\text{min})/p_T} \!\frac{du}{u^4\sqrt{u^2-1}} \\
     &+\int_{p_T/T(\tau_\text{min})}^{p_T/T(\tau_\text{max})} dz  z^4   \bar{C}_{\gamma}[f](z)  \int_{1}^{z T(\tau_\text{min})/p_T}\!\!\! \frac{du}{u^4 \sqrt{u^2-1}} \Big]. \nonumber 
\end{align}
The $u$-integrals can be done analytically and we get the final result for the photon spectra in ideal Bjorken expansion

\begin{align}\label{eq:idealfinite}
     &\frac{dN_\gamma}{d^2\xT d^2\pT dy} =\frac{\nu_\gamma}{(2\pi)^3} \frac{(\eta/s)^2}{\bar{p}_T^4}4\Big [ \\
     \quad&\int_{\sqrt{4 \pi \tilde w_\text{min}} \bar{p}_T}^\infty \hspace{-0.8cm} dz  z^4 \,  \bar{C}_{\gamma}[f](z)   (1-\frac{4\pi \tilde{w}_\text{min} \bar{p}_T^2}{z^2} )^{\frac{1}{2}}(1+\frac{4 \pi \tilde{w}_\text{min} \bar{p}_T^2}{2 z^2 })\nonumber \\
     \quad-&\int_{\sqrt{4 \pi\tilde w_\text{max}} \bar{p}_T}^\infty \hspace{-0.8cm} dz  z^4 \, \bar{C}_{\gamma}[f](z) (1-\frac{4\pi \tilde{w}_\text{max} \bar{p}_T^2}{z^2} )^{\frac{1}{2}}(1+\frac{4\pi \tilde{w}_\text{max} \bar{p}_T^2}{2 z^2 })\Big].\nonumber
\end{align}
Importantly, in the limit of infinite expansion, the spectrum is exactly a power law
\begin{align}
    \frac{dN_\gamma}{d^2\xT d^2\pT dy}   &\stackrel{\tilde{w}_\text{max}\to \infty}{\underset{\tilde{w}_\text{min}\to 0}{=}}\frac{\nu_\gamma}{(2\pi)^3} \frac{(\eta/s)^2}{\bar{p}_T^4}\tilde{C}_\gamma^\text{ideal}\label{eq:ideal}
\end{align}
where the normalization constant $\tilde{C}_\gamma^\text{ideal}$ is given by Eq.~(\ref{eq:CIdealDef}). Together with other constants from our simulations, the numerical values of $\tilde{C}_\gamma^\text{ideal}$ in leading order QCD kinetic theory for different couplings are listed in \cref{tab:CIdeal}.
\begin{table}[h]
    \centering
    \begin{tabular}{c|c|c|c}
        $\lambda$ & 5 & 10 & 20 \\
        \hline
        $\eta/s$ & 2.716 & 0.994 & 0.385 \\
        \hline
        $(\tau^{1/3} T)_\infty/Q_s^{2/3}$ & 0.799 & 0.612 & 0.474 \\
        \hline
        $\tilde{C}_\gamma^\text{ideal}/\sum_s q_s^2$ & 0.580 & 1.019 & 1.791 \\
        \hline
        $\tilde{w}_{\text{min}}$ & 0.015 & 0.035 & 0.077
    \end{tabular}
    \caption{Shear viscosity to entropy density ratio $\eta/s$, asymptotic temperature $(\tau^{1/3} T)_{\rm \infty}/Q_s^{2/3}$, photon production constant $\tilde{C}_\gamma^\text{ideal}/\sum_s q_s^2$ and initialization time of the simulation $\tilde{w}_{\rm min}$ for different coupling strengths $\lambda$.}
    \label{tab:CIdeal}
\end{table}

\section{Photon production from non-equilibrium QGP evolution}\label{sec:results} 

We will now present our simulation results for the photon emission from kinetically and chemically equilibrating QGP. We start the kinetic evolution at $\tau_0=1/Q_s$ with CGC-inspired gluon-dominated initial state. We use the same parametrization as in previous studies for the gluon distribution~\cite{Kurkela:2015qoa,Kurkela:2018oqw,Du:2020dvp} and set the initial fermion distribution to zero
\begin{subequations}
    \begin{align}
        f_g(\tau_0,p_T,p_z) &= \frac{2A}{\lambda} \frac{Q_0}{\sqrt{p_T^2 + \xi_0^2 p_z^2}} e^{-\frac{2}{3} \frac{p_T^2 + \xi_0^2 p_z^2}{Q_0^2}}\, ,\\
      f_q(\tau_0,p_T,p_z) &= 0 \, ,
    \end{align}
\end{subequations}
where $A=5.34$, $Q_0=1.8Q_s$, $\xi_0=10$ and $\lambda = g^2N_c = 4 \pi N_C \alpha_S$. How such system approaches chemical and kinetic equilibrium has been studied in detail in previous works~\cite{Kurkela:2018oqw,Du:2020dvp}. 

\subsection{Instantaneous rates}
\begin{figure*}
\centering
\begin{subfigure}[b]{0.24\textwidth}
\includegraphics[width=\textwidth]{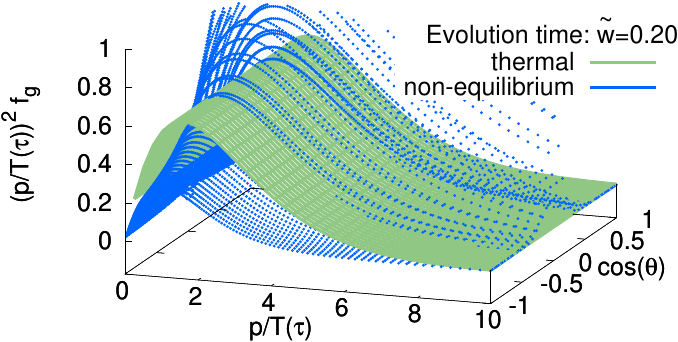}
\includegraphics[width=\textwidth]{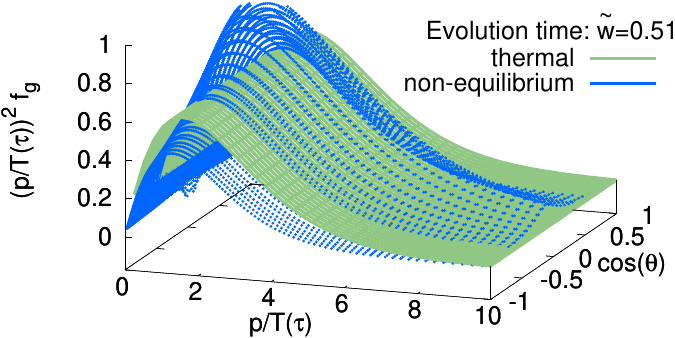}

\includegraphics[width=\textwidth]{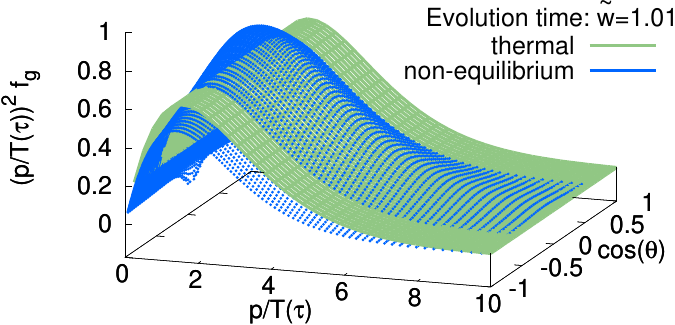}
    \caption{Gluon evolution}
    \label{fig:gluon}
\end{subfigure}
\begin{subfigure}[b]{0.24\textwidth}

\includegraphics[width=\textwidth]{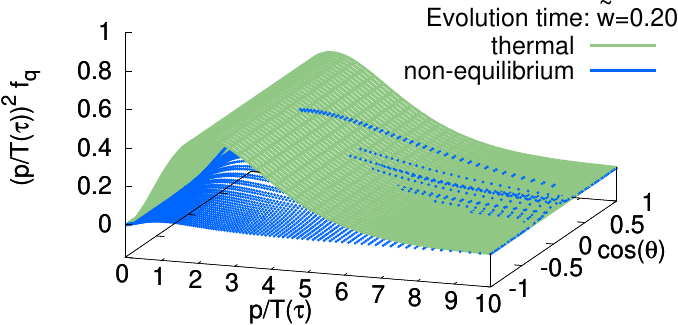}

\includegraphics[width=\textwidth]{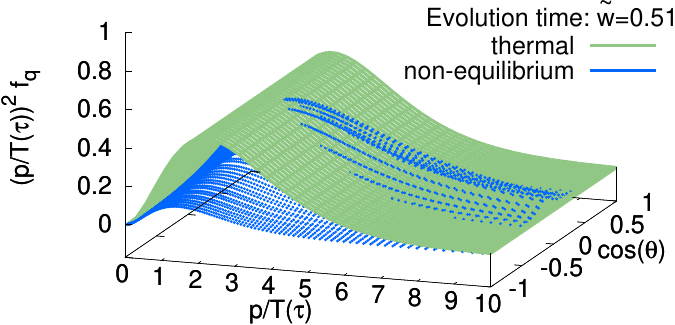}

\includegraphics[width=\textwidth]{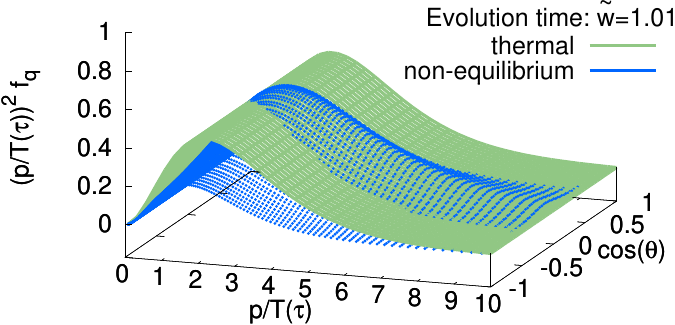}
    \caption{Quark evolution}
    \label{fig:quark}
\end{subfigure}
\begin{subfigure}[b]{0.24\textwidth}
\includegraphics[width=\textwidth]{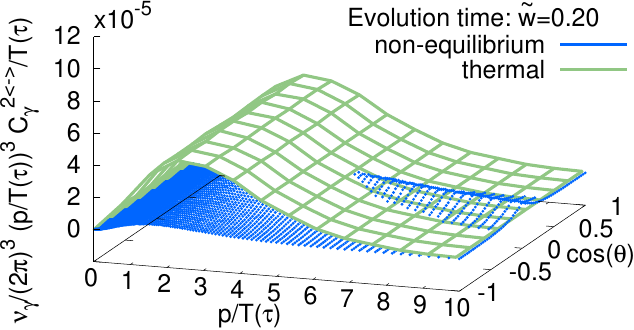}

\includegraphics[width=\textwidth]{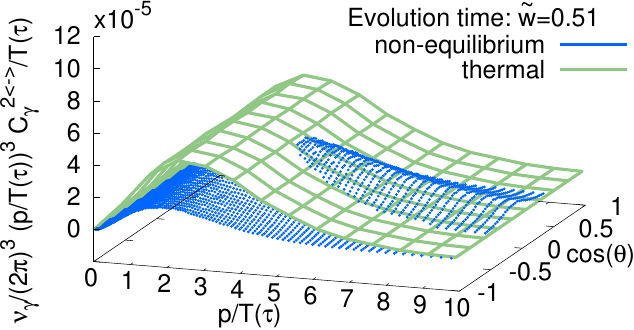}

\includegraphics[width=\textwidth]{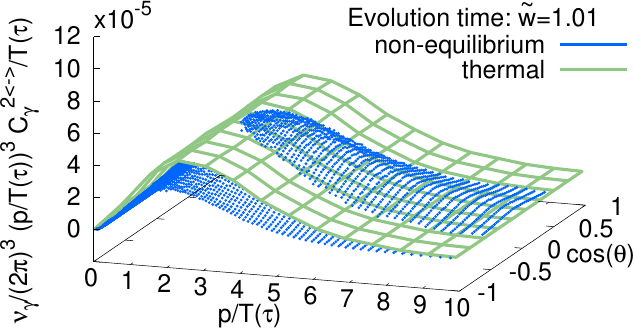}
    \caption{Elastic processes}
    \label{fig:el}
\end{subfigure}
\begin{subfigure}[b]{0.24\textwidth}
\includegraphics[width=\textwidth]{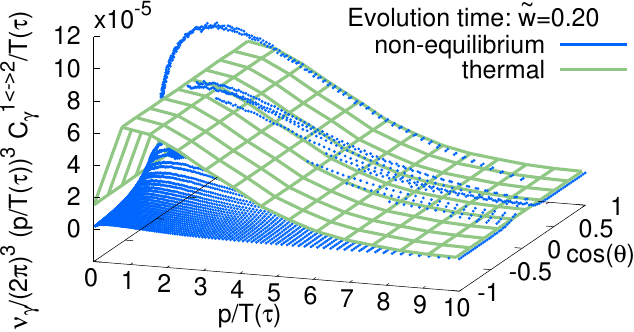}

\includegraphics[width=\textwidth]{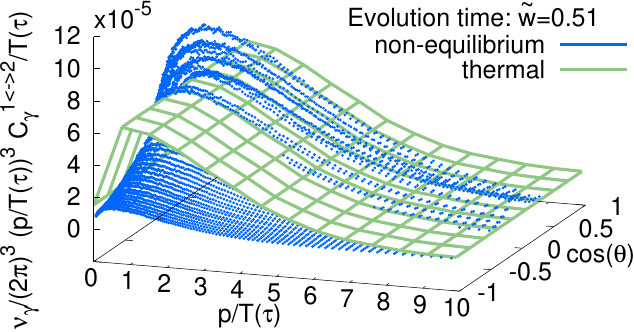}

\includegraphics[width=\textwidth]{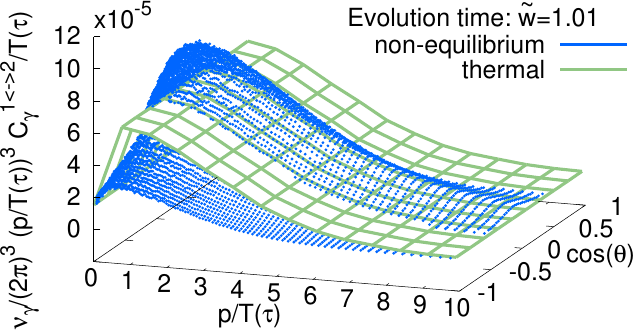}
    \caption{Inelastic processes}
    \label{fig:inel}
\end{subfigure}

\caption{\label{fig:DistributionFunctions} Thermalization of distribution functions for (a) gluons and (b) quarks in QCD kinetic theory at leading order and the corresponding leading order photon emission rates from (c) elastic and (d) inelastic processes.
Blue points correspond to non-equilibrium distribution functions or rates, while green lines show expected thermal equilibrium distribution resp. rate. Each row corresponds to a different scaled time $\tilde{w}=0.2,0.5,1.0$. Results are for $\lambda=10$.
}
\end{figure*}

In \cref{fig:gluon,fig:quark} we show the quark and gluon distributions for different times as a function of momentum $p/T_\text{eff}(\tau)$ and longitudinal momentum fraction $\cos\theta=p^z/p$ for $\lambda = 10$ ($\alpha_s=0.27$). The blue points are the non-equilibrium distribution functions and green lines correspond to the usual Fermi-Dirac distribution for quarks and Bose-Einstein distribution for gluons at temperature $T=T_\text{eff}(\tau)$, see \cref{eq:landau}. At very early times the $f_q$ is essentially zero, while $f_g$ is highly peaked at $\cos\theta=0$ due to the initial momentum anisotropy. As the system evolves, quarks are produced from gluon fusion $gg\to q\bar q$ and gluon splitting $g\to q\bar q$, and they also tend
to peak around $\cos\theta=0$. However as the longitudinal expansion becomes less prominent with time, the distributions become more isotropic (flatter in $\cos \theta$), but for $\tilde w =1$, the time around which hydrodynamization starts, the gluons are still over-populated and quarks are under-populated with respect to their thermal distributions. Eventually, for large $\tilde{w} > 2$ (data not shown) both distribution functions smoothly approach the respective thermal distributions.

In \cref{fig:el,fig:inel} we show the photon production rate split up into elastic and inelastic channels according to 
\begin{align}
   \frac{dN_\gamma}{d^4X d^3\pp} = \frac{\nu_\gamma}{(2\pi)^3} C_{\gamma}^{1\leftrightarrow 2}[f](x,\pp)+\frac{\nu_\gamma}{(2\pi)^3} C_{\gamma}^{2\leftrightarrow 2}[f](x,\pp) \, .
\end{align}
Once again blue curves show the non-equilibrium rate, while the corresponding thermal production rates are shown in green.\footnote{We compute the thermal rates by performing precisely the same calculation of the collision integrals, but using thermal distributions with the same total energy density (green lines in \cref{fig:gluon,fig:quark}).}
%
%
%
At early times all processes are essentially zero because there are no quarks in the plasma at the beginning. In the elastic channel shown in \cref{fig:el}, the Compton scattering and elastic pair annihilation make comparable contributions. The rate is significantly suppressed compared to thermal, because of suppressed quark distributions. We also see the slight peak at $\cos \theta= 0$, where the quark distribution is the highest.
In \cref{fig:inel} we show photon production in inelastic processes, where Bremsstrahlung dominates over the inelastic pair annihilation, because there are not many quarks to annihilate. We see a pronounced peak at $\cos\theta=0$. For larger couplings in thermal equilibrium (see \cref{fig:AMYrates}) the inelastic processes are more dominant than elastic photon production. We see that the same is true out of equilibrium for $\lambda=10$. Evolving the system further in time shows that we start to recover the thermal production rates at times $\tilde{w} \gtrsim 2$ (data not shown).

\begin{figure}
\centering
\includegraphics[width=0.49\textwidth]{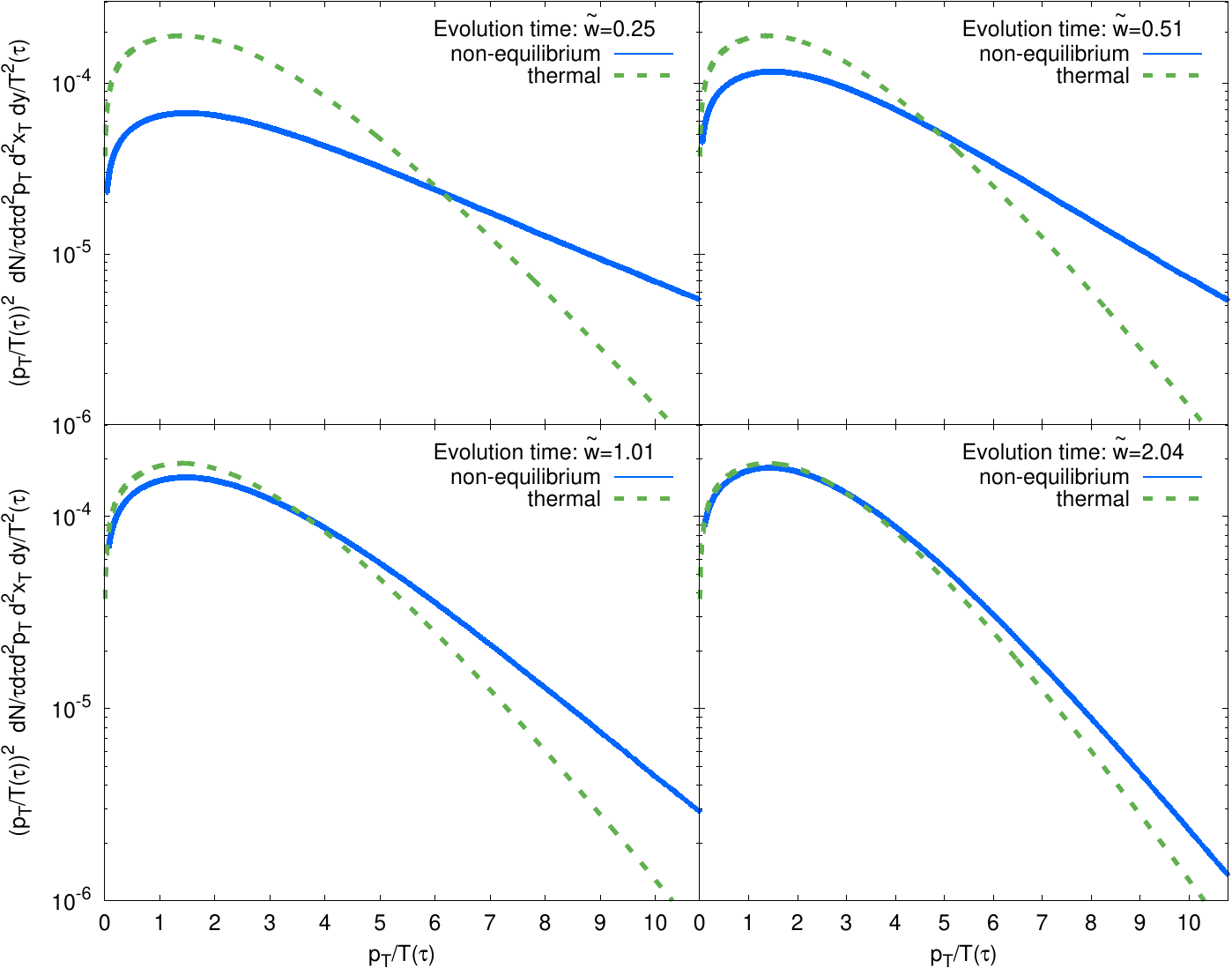}
\caption{\label{fig:InstantaneousRate1D} The sum of all four contributions to the photon production rate at leading order as a function of $p_T/T_\text{eff}(\tau)$ (subscript \textsl{eff} dropped for better readability). Blue solid curves correspond to non-equilibrium production, while green dashed lines represent the thermal contribution. Different panels correspond to different evolution times $\tilde{w}=0.25,0.5,1.0,2.0$. Results are for $\lambda=10$.}
\end{figure}

For the comparison to experimental measurements, we are interested in the $p_T$-spectrum. The corresponding production rate ${dN_\gamma}/{(\tau d\tau d^2\xT d^2\pT dy)}$ is given by the $p^z$ integral of the collision kernel in the fluid rest frame (see \cref{eq:spectra2}). In \cref{fig:InstantaneousRate1D} we show the results for $\lambda=10$ as a function of $p_T/T_\text{eff}(\tau)$ for different values of scaled time $\tilde w$. The non-equilibrium results are presented by blue solid lines, while the thermal rate is presented by green dashed lines.
At early times the production rate is well below the thermal rate at low momentum. However the non-equilibrium spectrum is much harder. As we scale the momentum with $T(\tau)=T_\text{eff}(\tau)$, which is the largest at early times, we emphasize that the low $p_T/T(\tau)$ regime is produced earlier than hard regime. This is manifested in the fact that the thermal production rate is recovered in the soft regime first. At later times $\tilde w>1$ the EKT spectrum also approaches the equilibrium at higher momentum although it still slightly overshoots the thermal spectrum.


\begin{figure}
\centering
\includegraphics[width=0.49\textwidth]{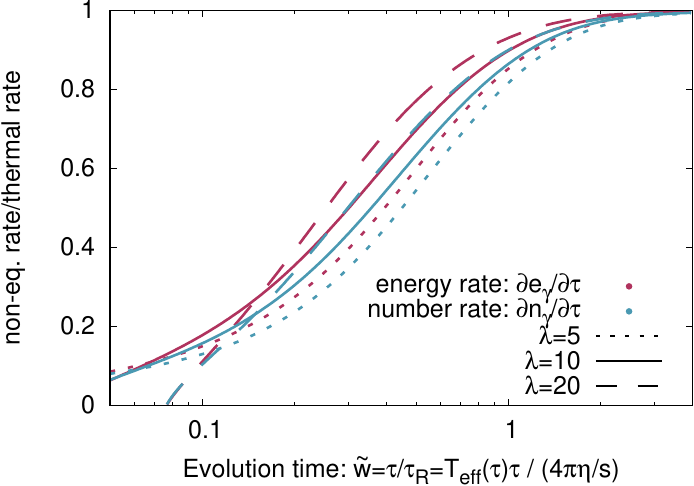}
\caption{\label{fig:EMLoss} Non-equilibrium photon production rate $\partial n_{\gamma}/\partial \tau$ and energy loss due to photon emission $\partial e_{\gamma}/\partial \tau$ compared to the corresponding thermal rates as a function of the evolution time $\Tilde{w}$. The red curves correspond to $\partial e_{\gamma}/\partial \tau$, while blue corresponds to $\partial n_{\gamma}/\partial \tau$. Different dash-types represent different coupling strengths.}
\end{figure}

Finally, we look at the fully integrated photon production rate ${\partial n_\gamma}/{\partial \tau}$ and the energy loss rate ${\partial e_\gamma}/{\partial \tau}$ due to photon production, which can be obtained as
\begin{subequations}
    \begin{align}
        \frac{\partial e_\gamma}{\partial \tau}=\frac{dE_\gamma}{d^4 X} &= \nu_\gamma \int \frac{d^3\vec{p}}{(2\pi)^3} p  C_\gamma(\tau,\pp) \, , \\
         \frac{\partial n_\gamma}{\partial \tau}=\frac{dN_\gamma}{d^4 X} &= \nu_\gamma\int \frac{d^3\vec{p}}{(2\pi)^3} ~C_\gamma(\tau,\pp) \, .
    \end{align}
\end{subequations}
\cref{fig:EMLoss} shows the photon energy and number rates compared to the thermal rates as a function of $\tilde{w}$ for different couplings $\lambda=5,10,20$. We see that the relative modification with respect to the thermal rates is very large at early times $\tilde{w} \ll 1 $ and quite similar for different couplings, supporting our argument of a universal modification of thermal rates. It is also important to point out, that for $\tilde w>1 $ corresponding to the typical time scale for the onset of hydrodynamic behavior, the non-equilibrium photon production rate converges to the expected equilibrium photon rate up to small residual non-equilibrium correction.

\subsection{Scaling of the time-integrated photon spectrum}\label{sec:ResultsYields}

\begin{figure}
\centering
\includegraphics[width=0.49\textwidth]{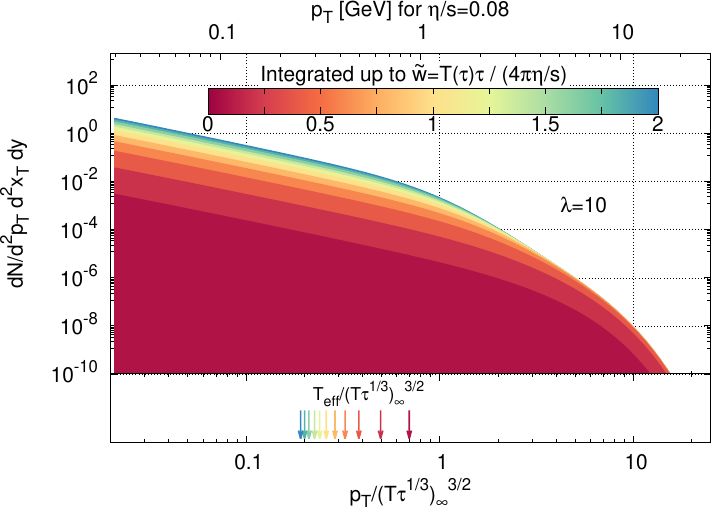}
\caption{\label{fig:Rainbow} Time integrated photon production rate as a function of $p_T/(T\tau^{1/3})_\infty^{3/2}$. Different colors represent integration up to a specific time $\Tilde{w}$ as indicated by the color box. At the bottom the $T_\text{eff}(\tau)$ corresponding to the different times are indicated. Top axis is the transverse momentum in physical units for typical $(T\tau^{1/3})_\infty^{3/2} = 0.402\, \text{GeV}$ (see discussion around \cref{eq:Tinfty}) and $\eta/s=0.08$. Results are for $\lambda=10$.}
\end{figure}

The total contribution from pre-equilibrium photon production is given by the time-integrated spectrum \cref{eq:spectra2}. Our goal in this section is to verify that the photon spectrum from the full QCD kinetic simulations satisfies the scaling formula \cref{eq:photonattractor} derived in \cref{sec:NonEqScalingOfYield} and extract the scaling curve. Therefore we will use the scaled transverse momentum variable $\bar{\mathbf{p}}_T ={(\eta/s)^{1/2}} \pT/{(\tau^{1/3} T)^{3/2}_\infty}$ to compare simulations with different values of the coupling constant $\lambda$ corresponding to different values of the shear viscosity to entropy density ratio $\eta/s$.

\begin{figure*}
\begin{center}
\centering
\begin{subfigure}[b]{0.45\textwidth}
\includegraphics[width=\textwidth]{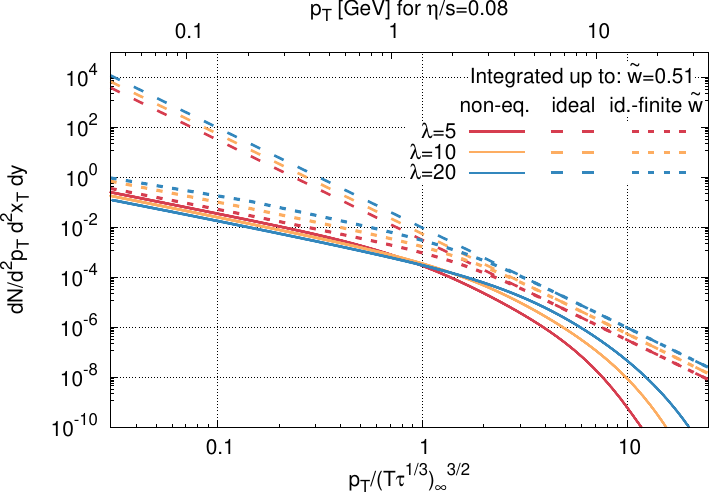}
\vspace{0.5pt}

\includegraphics[width=\textwidth]{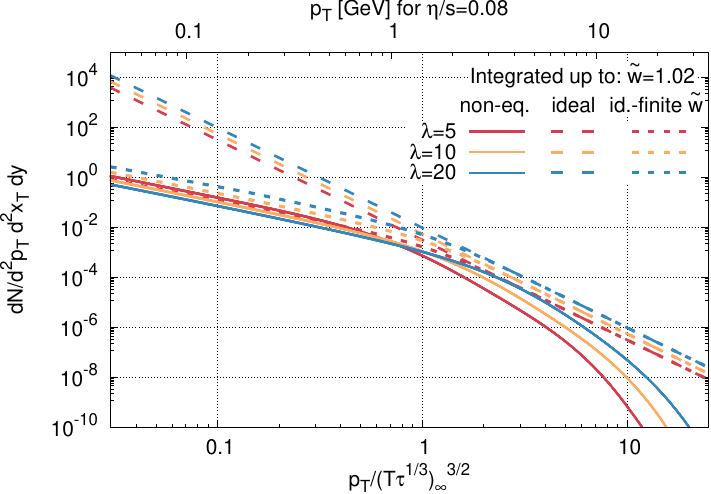}
\caption{Photon spectra}\label{fig:pTspectruma}
\end{subfigure}
\begin{subfigure}[b]{0.45\textwidth}
\includegraphics[width=0.89\textwidth]{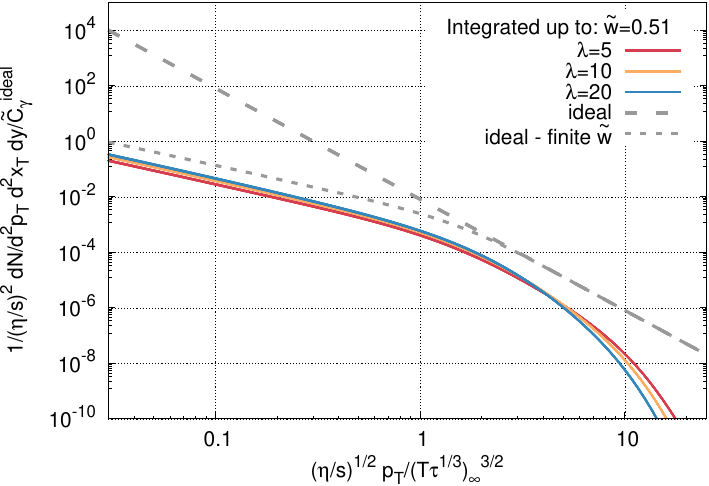}
\vspace{1.cm}

\includegraphics[width=0.89\textwidth]{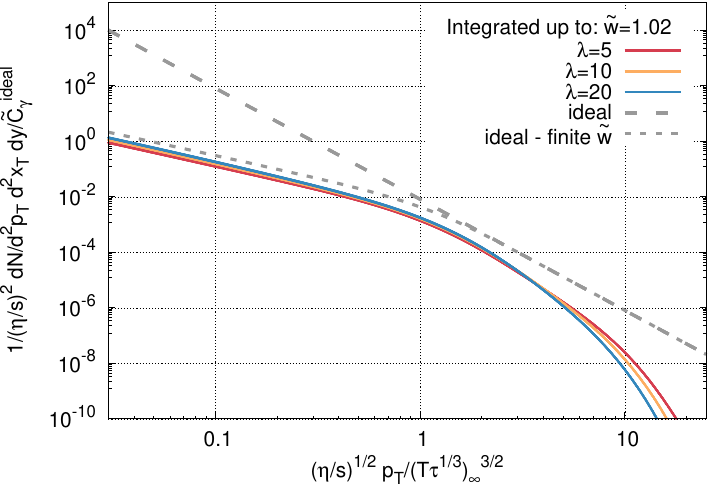}
\caption{Universal scaling curve}\label{fig:pTspectrumb}
\end{subfigure}
\caption{\label{fig:pTSpectrum} Spectrum of photons produced during the pre-equilibrium stage as a function of (a) $p_T/(T\tau^{1/3})_\infty^{3/2}$ and (b) $(\eta/s)^{1/2}p_T/(T\tau^{1/3})_\infty^{3/2}$. Different colors correspond to different coupling strengths $\lambda=5,10,20$ and different panels correspond to different final integration times $\tilde w=0.5,1.0$. The spectra in the right panels are divided by $(\eta/s)^2 ~\tilde{C}_\gamma^\text{ideal}$.}
\end{center}
\end{figure*}
In QCD kinetic simulations we extract the specific shear viscosity $\eta/s$ by fitting the late time behavior of the longitudinal pressure $P_L\equiv \tau^2 T^{\zeta\zeta}$ to the first order viscous hydrodynamics expectation for Bjorken expansion~\cite{Kurkela:2018vqr,Du:2022bel}
\begin{equation}
\frac{P_L}{e}\approx \frac{1}{3}-\frac{16}{9}\frac{\eta/s}{\tau T_\text{eff}(\tau)}.
\end{equation}
Similarly, we extract the asymptotic constant $(\tau^{1/3} T)_\infty$ by fitting the late time behavior of the effective temperature $T_\text{eff}(\tau)$ to the exact result in first order viscous hydrodynamics
\begin{equation}
   \left(\tau^{1/3}T \right)_\infty= \tau^{1/3}\,\left(T_{\text{hydro}}(\tau) + \frac{2}{3} \frac{\eta/s}{\tau}\right).\label{eq:hydromatch}
\end{equation}
We summarized the extracted values in \cref{tab:CIdeal}.

Of course, for comparisons with experimental data, it is important to estimate the typical value of $\scriptstyle\left(\tau^{1/3}T \right)_\infty$ in heavy ion collisions.
This is done by considering the average entropy per unit rapidity $dS/d\zeta = A_\perp \tau s(T)$, where $A_\perp$ is the transverse area of the collision. At sufficiently late times after the collision, so that the system is close to local thermal equilibrium, but early enough that the transverse expansion can be neglected, the entropy density per rapidity is constant $\tau s(T)= \frac{4\pi^2}{90}\nu_\text{eff}\left(\tau T^3 \right)_\infty$. Neglecting the entropy production during the transverse expansion, the same $dS/d\zeta$ entropy per rapidity is converted to hadrons on the freeze-out surface. Using the known average entropy per charged particle $S/N_\text{ch}\approx 7.5$~\cite{Hanus:2019fnc} we relate the charged particle multiplicity and entropy per rapidity as $\frac{dN_\text{ch}}{d\zeta} = \frac{N_\text{ch}}{S} \frac{dS}{d\zeta}$. We estimate the transverse area $A_\perp=138\,\text{fm}^2$. Using $\frac{dN_\text{ch}}{d\zeta} \approx 1600$~\cite{ALICE:2010mlf} for central Pb-Pb collisions at LHC energies we then find the typical value of $\left(\tau T^3\right)_\infty$ to be
\begin{align}
    \left(\tau T^3\right)_\infty &= 4.173\, \text{fm}^{-2} ~\left( \frac{dN_\text{ch}/d\zeta}{1600} \right) \left( \frac{S/N_\text{ch}}{7.5} \right) \nonumber \\
    &\times \left( \frac{A_\perp}{138\text{fm}^2} \right)^{-1} \left( \frac{\nu_\text{eff}}{\nu_g + 2 N_f ~\frac{7}{8} ~\nu_q} \right)^{-1} \, ,\label{eq:Tinfty}
\end{align}
which corresponds to
\begin{align}
    (T\tau^{1/3})_\infty^{3/2} = 0.402\, \text{GeV} \, .\label{eq:typicalT}
\end{align}
With this we can convert the scaled transverse momentum into physical units.

The results of QCD kinetic theory simulations for time-integrated production rate for $\lambda =10$ are shown in \cref{fig:Rainbow}. The different colors indicate up to which scaled-time $\tilde{w}$ the photon spectrum is integrated. The arrows at the bottom indicate the effective temperature $T_\text{eff}$ at the corresponding time. On the upper $x$-axis we show the momentum in physical units for the typical value of $ (T\tau^{1/3})_\infty^{3/2}$, see \cref{eq:typicalT}.
At early times the effective temperature is high and correspondingly the photon spectrum extends to much higher $p_T$ in absolute units. This regime is mainly produced up to times $\tilde{w}\sim 0.75$. At lower momentum, we find a power law behavior. Although the slope stays the same for all times, the later time contributions are added at lower momentum. Therefore the spectrum at intermediate momenta gains a different power law behavior and approaches the ideal Bjorken evolution (see \cref{eq:ideal}). At later times (from $\tilde{w}>1$) the spectrum is populated mostly at momentum below $p_T<0.4\,\text{GeV}$.




In \cref{fig:pTspectruma} we show the integrated spectra for different values of the coupling constant $\lambda=5,10,20$ and for different final times $\tilde w = 0.5,1.0$.
We compare QCD kinetic theory results (solid lines) to an idealized Bjorken expansion, where the system is assumed to be in thermal equilibrium throughout the entire evolution, which is shown by dashed lines (see \cref{sec:Bjorken}). In accordance with the discussion in \cref{sec:Bjorken}, for an infinite expansion, this 'ideal' spectrum is a simple power-law (see \cref{eq:ideal}). As dotted lines we show the thermal production rates for finite $\tilde{w}$ cut-off (see \cref{eq:idealfinite}), where we employed $\tilde{w}_\text{min}=0$. For high $p_T$ they collapse on the appropriate 'ideal' curves, which is due to setting $\tilde{w}_\text{min}=0$. At lower $p_T$ the curves with finite integration time deviate from the idealized curves as an effect of $\tilde{w}_\text{max}<\infty$.
We see that a larger coupling corresponds to a larger ideal photon spectrum throughout the whole range of $p_T$. Looking at the non-equilibrium contribution, we see that at high momentum, which is the regime produced the earliest in the evolution, more photons are produced the larger the coupling is. However, going down in momentum there is a point, where all three coupling coincide. For momentum below this point the hierarchy is reversed and for smaller couplings the production rate of photons is higher. As time evolves, the point were all three couplings agree, moves towards lower momentum and shifts from $p_T\approx 0.8$GeV for $\tilde w = 0.51$ towards $p_T\approx 0.6$GeV for $\tilde w = 1.02$. Overall the pre-equilibrium photon production is suppressed compared to the ideal thermal photon production, due to the suppression of quarks in the initial state.

In \cref{fig:pTspectrumb} we replot the same spectra using the predicted attractor scaling. Namely, we use ${(\eta/s)^{1/2}} \pT/{(\tau^{1/3} T)^{3/2}_\infty}$ as $x$-variable and divide the spectrum on the $y$-axis by $(\eta/s)^2 \tilde{C}_\gamma^\text{ideal}$. By construction, the ideal Bjorken expansion spectra collapse to the same curve (gray dashed) for all couplings. The same behavior is found for the finite $\tilde w_\text{max}$ curve (gray dotted). The QCD kinetic theory results (solid lines) also become much closer to each other, especially at later times. This is the manifestation of the predicted attractor scaling in \cref{eq:photonattractor}. Only in the high-$p_T$ sector, different couplings do not scale in the same way. This regime is produced at early times and is barely modified afterwards, such that a description in terms of an attractor solution is not appropriate at the time of production.

From \cref{fig:pTspectrumb} we can thus extract the universal scaling function for the integrated photon spectrum in non-equilibrium QGP expansion:
\begin{align}
    \mathcal{N}_\gamma \left(\Tilde{w},\textstyle \frac{\sqrt{\eta/s} ~p_T}{\left( T\tau^{1/3} \right)_\infty^{3/2}} \right) \equiv \frac{1}{(\eta/s)^2 ~\tilde{C}_\gamma^\text{ideal}} ~\frac{dN}{d^2\xT d^2\pT dy}   
    \label{eq:ScalingFunction}
\end{align}
Generally, the scaling curve for typical values of $\tilde{w}=1$, the timescale on which the system hydrodynamizes, can be characterized as follows. At high regime, $\frac{\sqrt{\eta/s} ~p_T}{\left( T\tau^{1/3} \right)_\infty^{3/2}} \gtrsim 4$, it shows a steep fall-off as this regime is sensitive to the earliest evolution times, where only few quarks are present and thus few photons are produced. In the intermediate momentum region $1\lesssim \frac{\sqrt{\eta/s} ~p_T}{\left( T\tau^{1/3} \right)_\infty^{3/2}} \lesssim 4$ the scaling curve for pre-equilibrium photon production shows the same power-law behavior as the ideal result, albeit with an overall suppression, which we attribute to the continued quark suppression during the pre-equilibrium evolution. Below $\frac{\sqrt{\eta/s} ~p_T}{\left( T\tau^{1/3} \right)_\infty^{3/2}} \lesssim 1$ the scaling curve flattens and approaches a different power law behavior than in the intermediate region as the softer photons tend to be predominantly produced at later times $\tilde{w}>1$. This is confirmed by the fact that the pre-equilibrium scaling curve shows essentially the same behavior as the ideal spectrum
for a finite evolution time $\tilde{w}_\text{max}$, once again with a slight suppression due to the chemical non-equilibrium conditions.

\begin{figure}
\begin{center}
\includegraphics[width=\columnwidth]{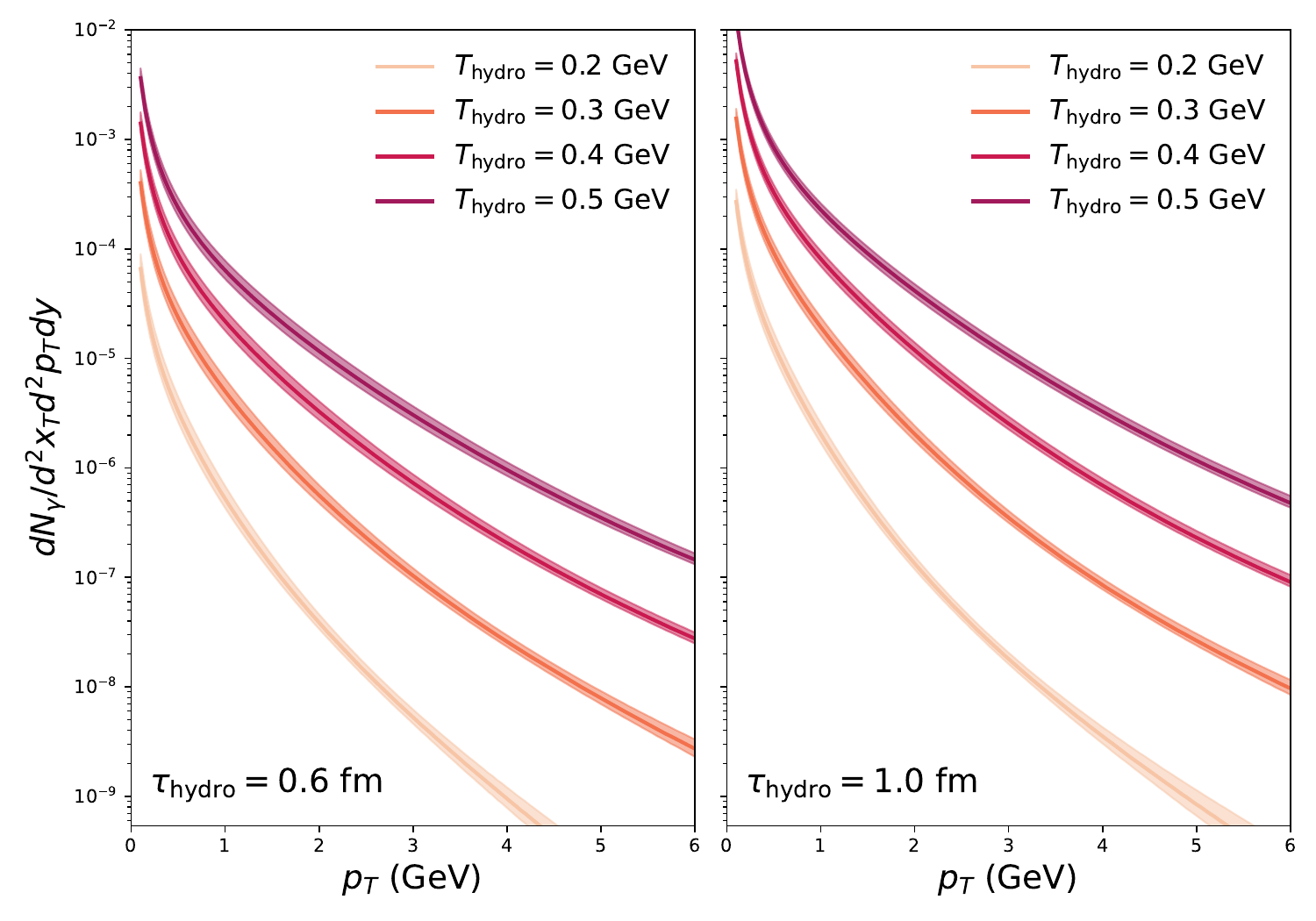}
\caption{Photon spectra from pre-equilibrium evolution with final temperature $T_\text{hydro}$ at $\tau_\text{hydro}$ for $\eta/s= 0.08-0.16$ variation shown as bands. We use $\tilde{C}_\gamma^\text{ideal}=0.573$ corresponding to AMY parameterization with $\lambda=10$ (see \cref{tab:CIdeal}).
The different panels correspond to $\tau_\text{hydro}=0.6$~fm (left) and $\tau_\text{hydro}=1.0$~fm (right).
\label{fig:TempDependence}}
\end{center}
\end{figure}

One of the remarkable consequences of the universality of the scaling result of \cref{eq:ScalingFunction} is the practical usefulness for phenomenology. Based on the scaling function in \cref{eq:ScalingFunction}, it is directly possible to compute the pre-equilibrium photon production for a given initial condition for a hydrodynamic simulation by matching the viscosity $\eta/s$ and the energy density at a matching time $\tau_\text{hydro}$. In \cref{fig:TempDependence} we show the photon spectra from the non-equilibrium rate that is integrated up to time $\tau_\text{hydro}$ at which $T_\text{eff}(\tau_\text{hydro})=T_\text{hydro}$. The constant $\scriptscriptstyle\left( T\tau^{1/3} \right)_\infty$ is obtained from \cref{eq:hydromatch} (assuming that entropy evolution is well described by 1st order hydrodynamic Bjorken expansion after $\tau_\text{hydro}$) and the scaled time at $\tau_\text{hydro}$ is simply $\tilde w_\text{hydro} =\tau_\text{hydro} T_\text{hydro}/(4\pi \eta/s)$. The two panels in \cref{fig:TempDependence} show the results for two different switching times $\tau_\text{hydro}=0.6$\,fm (left) and $\tau_\text{hydro}=1.0$\,fm (right), where for a fixed temperature $T_\text{hydro}$ a larger $\tau_\text{hydro}$ corresponds to a larger $\tilde{w}_\text{hydro}$ and larger $\scriptscriptstyle\left( T\tau^{1/3} \right)_\infty$. In each panel the photon spectra is shown for different temperatures $T_\text{hydro}$. Bands in \cref{fig:TempDependence} correspond to the results obtained for different values of $\eta/s$ in the range of $0.08-0.16$. Clearly, the first thing to notice is that for a fixed matching time $\tau_\text{hydro}$ fluid cells with higher final temperature $T_\text{hydro}$ will produce more photons in the non-equilibrium evolution. Conversely, photon production from regions with smaller local temperatures is strongly suppressed. Secondly a later initialization time also results in a larger pre-equilibrium photon production yield, in particular at low $p_T$ as these photons tend to be produced later over the course of the evolution. By looking at the relative size of the bands in \cref{fig:TempDependence}, one observes that the variation of the pre-equilibrium photon yield with changing $\eta/s$ is rather small, indicating that uncertainties in the photon yield that stem from the microscopic dynamics of thermalization are not too dramatic. Although this might be surprising at first, it should be expected considering the momentum range sampled in \cref{fig:TempDependence}. For the ideal spectrum in \cref{eq:ideal} we observe an exact cancellation of $\eta/s$ by $1/\bar{p}_T^4$. The reader can observe that the physical momentum in \cref{fig:TempDependence}, $p_T\sim 1-5$~GeV, corresponds to the universal momentum, $p_T(T\tau^{1/3})^{-3/2}\sim 1-10$.  In such range the pre-equilibrium spectrum is close to the ideal one and we see an (almost exact) compensation and consequently only a small $\eta/s$ dependence. Additionally, the spectrum does not depend on $\tilde{w}_\text{max}$, which would give another source of $\eta/s$ dependence, as \cref{fig:Rainbow} shows that this momentum range is actually produced on timescales, which are well below the cut-off in $\tilde{w}$ applied in our QCD kinetic simulations.




\section{Phenomenology of the pre-equilibrium photons}\label{sec:pheno} 
 \begin{figure*}
\begin{center}
\includegraphics[width=0.9\textwidth]{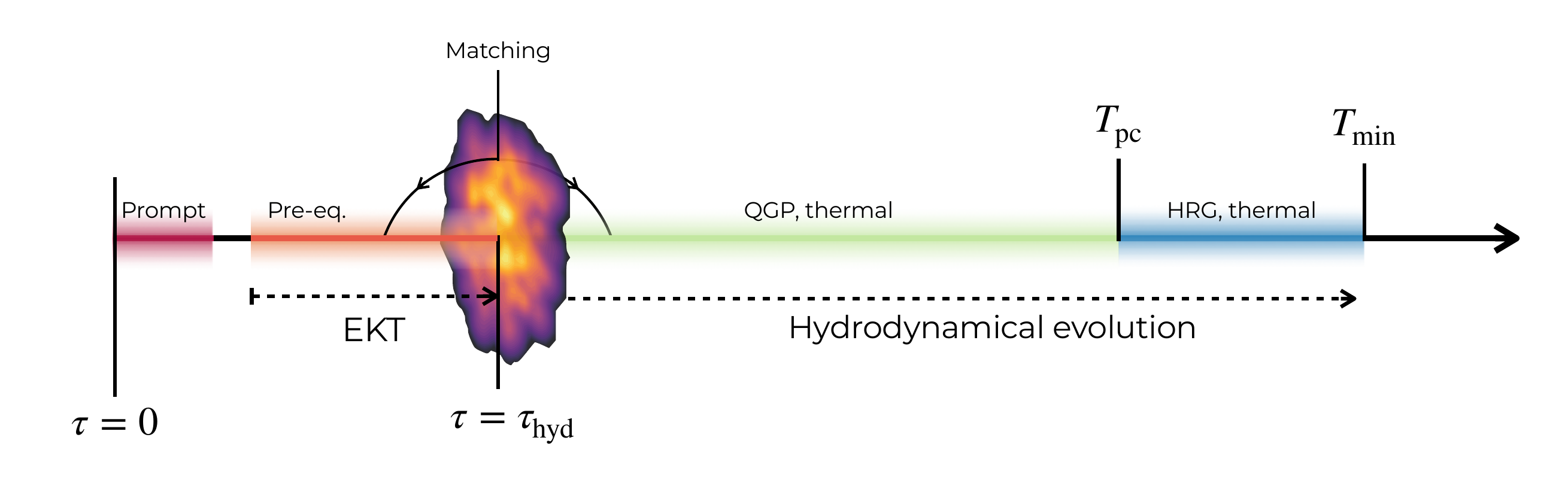}
\caption{\label{fig:Pheno} The schematic timeline for photon production in heavy-ion collisions. The prompt photons are produced in hard scatterings at the instant of the collision. The pre-equilibrium photons (this work) are radiated during the first $\sim 1\,\text{fm}$ of the evolution as the QGP is equilibrating. At $\tau_\text{hydro}$
the pre-equilibrium production in QCD kinetic theory is smoothly matched (locally in the transverse plane and event-by-event) to equilibrium (AMY) photon rates in the hydrodynamic phase.
%
%
The matching procedure is described in \cref{sec:matching}.
Afterwards, thermal photons from the QGP are produced by folding thermal AMY rates with the hydrodynamic evolution. As the system goes through the cross-over at $T_\text{pc}=160\,\text{MeV}$, we switch to hadronic rates, which are the source of emission until the end of the collision evolution at $T_\text{min}=120-140\,\text{MeV}$.}
\end{center}
\end{figure*}

We will now describe how our results for the pre-equilibrium photon production can be employed in phenomenological calculations of the direct photon production.
For phenomenological comparison to experimental photon data, the complete direct photon spectrum can be constructed as a sum of prompt and in-medium photons. The prompt photons are produced in hard scatterings at the initial instant of the collision, while the in-medium photons are a sum of the radiation coming from the deconfined medium (pre-equilibrium and equilibrium QGP) and photons produced by hadronic scatterings in an interacting hadron gas. A schematic timeline of different photon sources is summarized graphically in \cref{fig:Pheno}. Equilibrium QGP and hadronic radiation are relatively well-understood in-medium contributions. 
However, before this work the full computation of pre-equilibrium photons was lacking. In the next section we describe our procedure for computing pre-equilibrium photons and matching to the subsequent hydrodynamic evolution. For prompt, equilibrium QGP and hadron resonance gas contributions we follow Refs.~\cite{Garcia-Montero:2019kjk, Garcia-Montero:2019vju} and reproduce the details of the computation in \cref{sec:sourcesPheno}.



\subsection{Photons from the pre-equilibrium stage}
\label{sec:matching}
Pre-equilibrium photons are computed by matching the QCD kinetic theory evolution of the QGP to the initial conditions of the subsequent hydrodynamic evolution.
%
%
%
%
%
To compute the pre-equilibrium spectrum for the photons, the two main ingredients are the scaling formula, \cref{eq:ScalingFunction}, and the knowledge of the temperature profile at the start of the hydrodynamic evolution (see \cref{fig:Pheno}). Notice that the scaling formula readily provides the spectrum of all photons produced over the course of the pre-equilibirum evolution. Given an initial temperature profile $T(\tau_{\rm hydro},\xT)$ at the initialization time $\tau_{\rm hydro}$ of a hydrodynamic evolution with $\eta/s$, one can then extract the local value of the scaling variable $\tilde{w}(\xT)$, as well as the local energy scale $(T\tau^{1/3})_{\infty}(\xT)$ for each fluid cell $\xT$ to obtain the pre-equilibrium photon spectrum. The total spectrum is given by the integral in the transverse plane
\begin{equation}
    \frac{dN}{d^2\pT dy} = \left(\frac{\eta}{s}\right)^2  \Tilde{C}_{\gamma}^{\text{ideal}}\int d^2\xT \mathcal{N}_\gamma \left(\Tilde{w}(\xT),\textstyle \frac{\sqrt{\eta/s} ~p_T}{\left( T(\xT)\tau^{1/3} \right)_\infty^{3/2}} \right).
\label{eq:phenomastereq}
\end{equation}
The physical meaning of this is that we match the temperature in the fluid cell at hydrodynamic initialization time to the final state temperature in the kinetic theory evolution. By reading off the corresponding rescaled time $\tilde w(\xT)$ it is possible to look up the spectra of photons produced by such a cell during the preceding non-equilibrium evolution.



On a more pragmatic note, the $\eta/s$ used in the matching is taken as an ``external'' hydro parameter, and not the $\eta/s$ extracted from EKT for a given coupling constant $\lambda$. 
Here we follow the phenomenological practice~\cite{Kurkela:2018oqw,Kurkela:2018wud} of using the approximate independence of the coupling constant of the EKT evolution when expressed in scaled time $\tilde w$. The physical shear viscosity, which characterizes the realistic interaction strength of the QGP, is more reliably extracted from the hydrodynamic model to data comparisons~\cite{Nijs:2020ors,Nijs:2020roc} than perturbative computations due to large NLO corrections~\cite{Ghiglieri:2018dib}. For definiteness, we will use the value $\eta/s=0.08$, which is on the lower side of the extracted values.
Similarly, the normalization constant of thermal photon production $\tilde{C}_\gamma^\text{ideal}$ in \cref{eq:phenomastereq} can be taken as an external parameter. However, the NLO corrections to LO thermal photon production is small $\sim 20\%$ at $\alpha_s=0.3$~\cite{Ghiglieri:2013gia}. Therefore in the following we will take $\tilde{C}_\gamma^\text{ideal}=0.573$, which corresponds to AMY thermal rate for $\alpha_s=0.265$, which we used for thermal productions during the hydrodynamic evolution.




The quantity $(\tau^{1/3} T )_\infty$ sets the physical scale, c.f. \cref{eq:Tinfty}, and we will extract it using the exact formula at first viscous order
\begin{equation}
    (\tau^{1/3}T)_{\infty}(\xT) = \tau_{\rm hydro}^{1/3}\,\left(T_{\rm hydro}(\xT) + \frac{2}{3} \frac{\eta/s}{\tau_{\rm hydro}}\right).
\end{equation}
The next step is to find the rescaled time, $\tilde{w}$ which can be directly extracted as
\begin{equation}
   \tilde{w}(\xT)= \frac{\tau_{\rm hydro} T_{\text{hydro}}(\xT)}{4\pi \eta/s}.
   \label{eq:hydro_wtilde}
\end{equation}
%
Notice that in the above expressions, both the evolution time $\tilde{w}$ and the local energy scale $(\tau^{1/3}T)_{\infty}$ vary across the transverse plane, due to the $\xT$ dependence of the temperature profiles. We will use a constant $\eta/s$, but more generally transport coefficients in hydrodynamic simulations can be temperature dependent and therefore induce further $\xT$ dependence\footnote{In conformal theories, like massless QCD kinetic theory, $\eta/s$ is always temperature independent.}.

Additionally, by varying $\Tilde{w}_{\rm min}$, one can explore the sensitivity to the initial time for the photon spectrum. In practice this is performed by subtracting from the spectra up to the desired minimum universal time. Namely, this is achieved by substituting the scaling function in \eqref{eq:phenomastereq} for 
\small
\begin{equation}
\mathcal{N}_\gamma \left(\Tilde{w}(\xT),\textstyle \frac{\sqrt{\eta/s} ~p_T}{\left( T(\xT)\tau^{1/3} \right)_\infty^{3/2}} \right)-\mathcal{N}_\gamma \left(\Tilde{w}_{\rm min},\textstyle \frac{\sqrt{\eta/s} ~p_T}{\left( T(\xT)\tau^{1/3} \right)_\infty^{3/2}} \right)\,.
\end{equation}
\normalsize
In what follows, we apply this matching procedure to an event-by-event boost invariant (2+1D) hydrodynamical simulations. 
The realistic event-by-event temperature profiles in the transverse plane are taken from viscous hydrodynamic simulations tuned to experimental data.
We use the same 200 Pb-Pb events 
at $\sqrt{s_{\rm NN}} = 2.76\,\text{TeV}$ corresponding to the 0--20$\%$ centrality class as in Refs.~\cite{Garcia-Montero:2019kjk, Garcia-Montero:2019vju}.
Hydrodynamics is initialized at $\tau_{0}=0.6\,\text{fm}$ with the two-component Monte Carlo–Glauber model~\cite{Shen:2014lye} and evolved with the VISHNU package \cite{Shen:2014vra,VISHNU} using the default model parameters. In particular, we use the lattice equation of state\footnote{Lattice QCD simulations gives $\nu_\text{eff}\sim 40$ at $T\sim 400\,\text{MeV}$, which is smaller than $\nu_\text{eff}=47.5$ in ideal QGP. Therefore matching temperature in EKT and hydrodynamic simulations results a small discontinuity in energy density.}, a constant specific shear viscosity of $\eta/s=0.08$ and zero bulk viscosity. See Refs.~\cite{Garcia-Montero:2019kjk, Garcia-Montero:2019vju} for further details.

Finally, we note that the matching is done between transversely homogeneous EKT simulations and (2+1D) hydrodynamics at a constant $\tau_{\rm hydro}\geq \tau_0$. 
This matching time should be sufficiently small compared to the transverse system size $\sim 10\,\text{fm}$, such that we could neglect any transverse dynamics, e.g., transverse velocity profile. However, the matching time does not need to be constant, as in principle one only needs the initial hypersurface of the hydrodynamic simulation, which provides both temperature and proper time profiles.


\subsection{Comparison to experimental data}

\begin{figure}
\begin{center}
\includegraphics[width=0.48\textwidth]{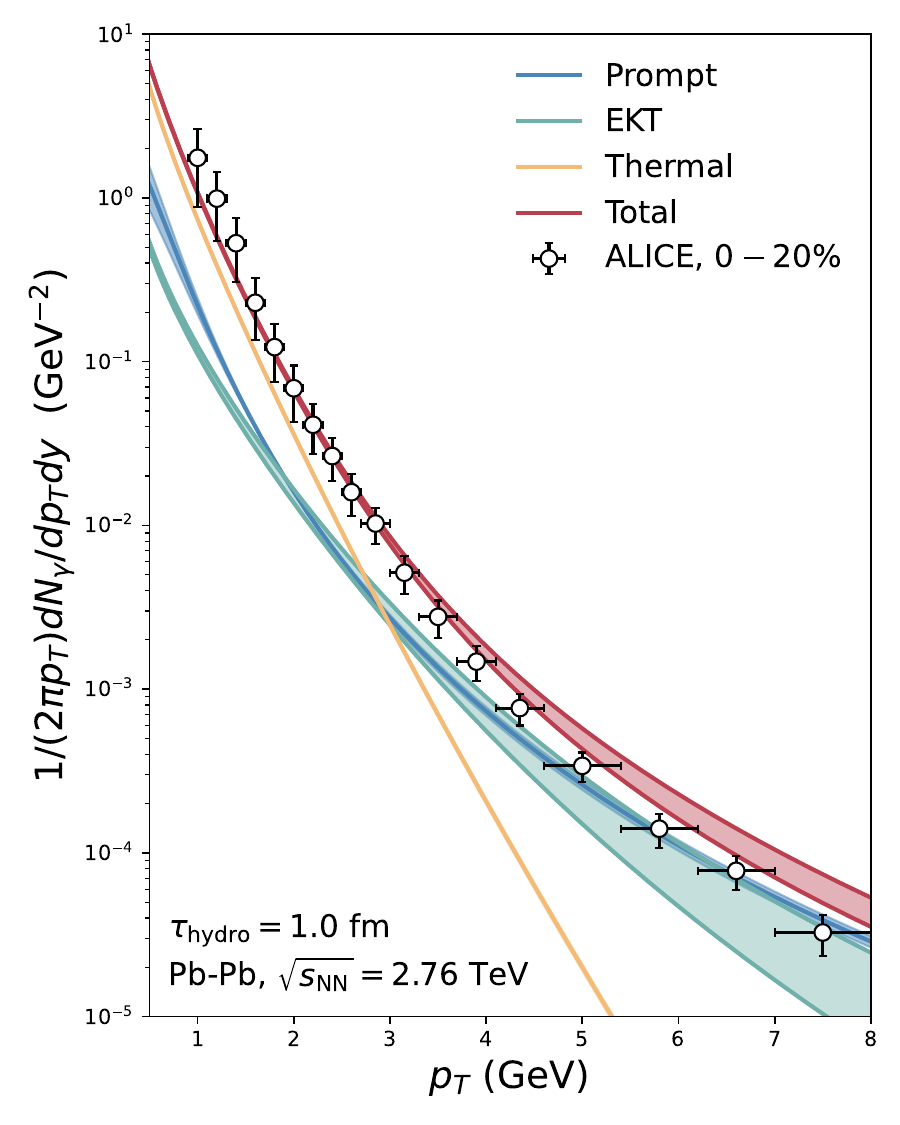}
\caption{\label{fig:pTYieldALICE} The differential photon spectra for $0-20\%$ centrality PbPb collisions at 2.76 TeV. The experimental ALICE results are shown as points \cite{ALICE:2015xmh,Sahlmuller:2015mfq}. The solid red lines show the total computed photon spectra, which consist of prompt (blue), pre-equilibrium (green) and thermal rates in the hydrodynamic phase (yellow).}
\end{center}
\end{figure}

Our main result for this section lies in \cref{fig:pTYieldALICE}, where we present our computation of the total photon yield in Pb-Pb collisions at $2.76$ TeV, for the $0-20\%$ centrality class. In \cref{fig:pTYieldALICE}, the multiple sources of photon production are presented such as the prompt (blue), the pre-equilibrium contribution (this work, green), and thermal photons from the QGP and the hadronic stage (yellow). We compare the summed total (red) to the measured ALICE photon yield~\cite{ALICE:2015xmh,Sahlmuller:2015mfq} and find a good agreement.

When computing the pre-equilibrium contribution, we employ the photon yields for $\lambda=10$, which corresponds to $\alpha_s=0.27$, while the details of the calculation of prompt photon production, as well as the production of photons from the thermal QGP and hadronic sources are identical to the procedure of~\cite{Garcia-Montero:2019kjk}, but nevertheless included in \cref{sec:sourcesPheno} for completeness. Additionally, we have included a variation of minimum universal time $\tilde{w}_\text{min}=0.1-0.5$ using the prescription described above.


Comparing the different sources, the thermal photon production is most prominent in the low momentum regime, whereas the EKT photons have the smallest contribution to the photon spectra in this region. Around $p_T\sim 2.5$~ GeV, the thermal contribution becomes the smallest one and drops off rapidly with increasing transverse momentum. Between $p_T\sim 2.5$~ GeV and $p_T\sim 7$~ GeV the EKT photons are comparable to the spectra coming from prompt photons. Above $p_T\sim 7$~ GeV the contribution from EKT falls off as well and the hard-momentum sector is mainly dominated by prompt photons.

When calculating the pre-equilibrium contribution in \cref{fig:pTYieldALICE}, we have chosen a matching time of $\tau_\text{hydro}=1.0\,\text{fm}$ (see \cref{sec:matching}). Nevertheless, this choice is not unique, as the formalism in this paper allows for a smooth transition between the early out-of-equilibrium stage and the hydrodynamical evolution if $\tilde{w}\gtrsim 1$~\cite{Kurkela:2018vqr}. We quantify the validity for the matching times range we have chosen to explore, $\tau_\text{hydro}=0.6-1.0\,\text{fm}$, by looking at the spatially averaged scaling time variable for the average event at the extrema in this range. We find
\begin{equation}
    \begin{split}
        \langle\tilde{w}(\tau_\text{hydro}=0.6\,{\rm fm})\rangle \approx 1.00\,,\\ 
        \langle\tilde{w}(\tau_\text{hydro}= 1\,{\rm fm})\rangle \approx 1.45\,,
    \end{split}
\end{equation}
for our fixed value of $\eta/s=0.08$, which is indeed in the regime where the non-equilibrium evolution of the QGP can be described hydrodynamically. 



\begin{figure}
\begin{center}
\includegraphics[width=0.45\textwidth]{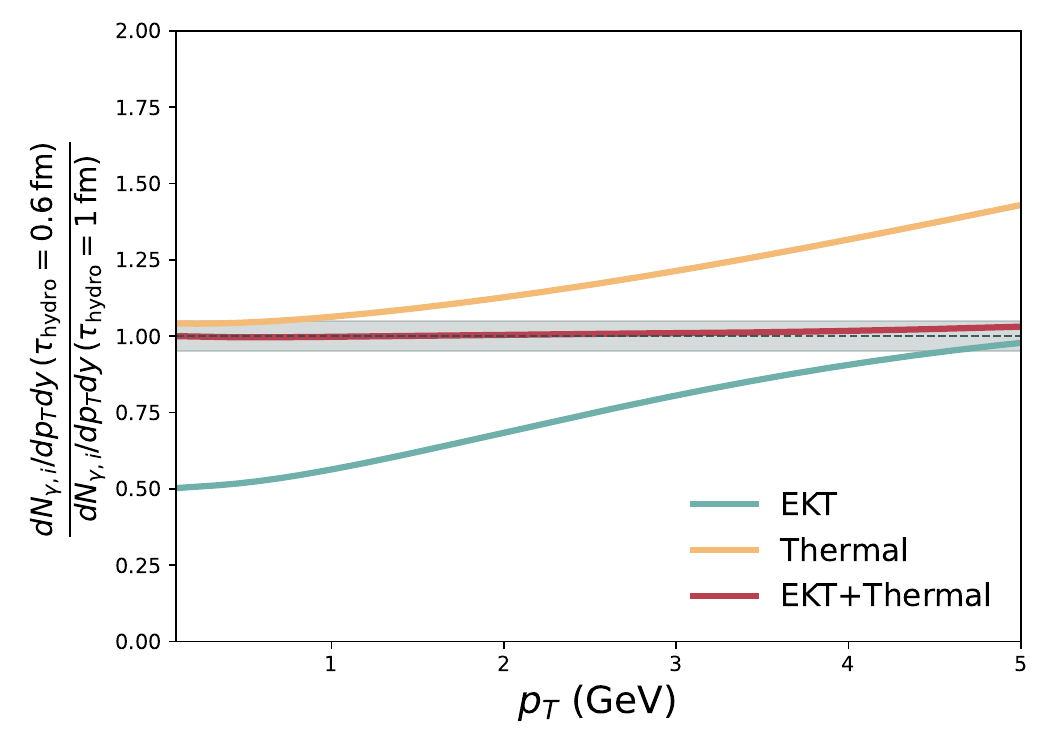}
\caption{The ratio of differential photon spectra in kinetic+hydrodynamic evolution computed for two matching times $\tau_\text{hydro}=0.6,1.0\,\text{fm}$. The yellow line is the ratio for thermal yields, green for EKT yields and red for the sum of EKT and thermal. The gray band correspond to changes of $5\%$.}
\label{fig:tHydroComp} 
\end{center}
\end{figure}

We explored the sensitivity of the resulting total yield to the variation of the matching time. In Fig. \ref{fig:tHydroComp} we show the ratio of the thermal and pre-equilibrium contributions where the matching is once performed at $\tau_{\rm hydro} = 0.6$~fm and once at $\tau_{\rm hydro} = 1$~fm. The EKT curve is shown in green and the thermal contribution in yellow, while red shows the sum of EKT and thermal. Notably, the EKT and thermal ratios both deviate signficantly from each other. For the thermal yield, the ratio is insensible to the switching at low momentum, while for higher $p_T$ an earlier matching time means a larger yield. For the EKT photon yield this behavior is somewhat reversed. At low momentum a later switching time means a higher yield, while the EKT production is insensible to $\tau_\text{hydro}$ at high $p_T$. Strikingly, the sum of both contributions is independent of the switching time ($<5\%$ variation). This allows for the matching to be smooth regardless of the matching time, as long as the relevant characteristic $\tilde{w}$ is in the regime where the non-equilibrium evolution of the QGP can be described hydrodynamically.

Finally, when varying the initial time, we seem to get better agreement for larger $\Tilde{w}_{\rm min}$ (lower end of the EKT band in \Cref{fig:pTYieldALICE}). The variation of the minimum time at which photons are produced is an interesting handle that needs to be explored further in future work. Nevertheless, we would like to note that, regarding the matching at large $p_T$, more uncertainties may come into play when calculating the prompt contribution as they are commonly shown in the literature. For example, while error is computed in this figure by accounting for a change of the renormalization scale of the prompt processes (see \Cref{sec:sourcesPheno}), the uncertainties of the parton distribution functions are not included in most of the literature. Additionally, the process is scaled using a simple MC-Glauber approach, which ignores the nuclear effects of the aforementioned distributions. These sources of uncertainties and novel parameterizations of nuclear effects need to be taken into account in the future to properly understand the photonic contributions at low and intermediate momenta.


\section{Summary and Outlook}\label{sec:conclusion}

We computed the photon emission in chemically equilibrating QGP using leading order QCD kinetic theory. For phenomenological relevant coupling strengths ($\lambda=10$, $\alpha_s=0.265$), we showed that inelastic processes (Bremsstrahlung, inelastic pair annihilation) dominate the production over the elastic ones (Compton scattering, elastic pair annihilation). Especially Bremsstrahlung off quarks and antiquarks dominates all other four processes for the three couplings tested here. At early times the rates are suppressed significantly since the system is gluon dominated and few quarks are produced yet. Evolving in time the system cools down and the non-equilbrium rates smoothly approach their counterparts in thermal equilibrium.

In addition we showed that photons with high $p_T$ are predominantly produced at the earliest stages of heavy-ion collisions. In this regime we see a steep fall-off compared to photon production in an idealized expansion due to a strong quark suppression at this stage of the evolution. Conversely, the low-$p_T$ regime is produced at late times, where the photon spectra shows a characteristic power law behavior. The slope of the curve stays the same for all times although for later times the total contribution in this regime increases. In between an intermediate regime is established, which exhibits a different power law behavior.

Assuming that the pre-equilibrium evolution of the QGP can be described in terms of a single scaled time variable $\tilde{w}$, we showed that the photon spectra satisfies a simple scaling formula where the momentum is scaled by ${(\eta/s)^{1/2}} \pT/{(\tau^{1/3} T)^{3/2}_\infty}$ with an additional overall normalization of $1/(\eta/s)^2\tilde{C}^\text{ideal}_\gamma$ to the photon spectra. Here $\tilde{C}^\text{ideal}_\gamma$ is a constant coming from the thermal photon spectra, if the system is assumed to be in equilibrium for all times. 

The benefit of this universality is that our results can be applied to estimate the pre-equilibrium photon production in realistic simulations of heavy ion collisions. We applied our formula in event-by-event simulations and found a smooth matching to hydrodynamic photon yields. Compared to different sources of photons during a heavy-ion collision, the pre-equilibrium yield is small and is dominated by thermal contributions for low momenta and by prompt photons for the whole considered range of momenta. Nevertheless above $p_T\sim2.5$GeV, the  contribution from pre-equilibrium exceeds the thermal photon production in the hydrodynamic phase and is almost of the same order as the prompt contribution.

We implemented our formalism into the initial state framework \kompost{} \cite{KoMPoST,Kurkela:2018vqr}. This version, called Shiny\kompost{}, allows the computation of pre-equilibrium photon spectra from the energy-momentum tensor profile at the hydrodynamic starting time $\tau_\text{hydro}$. Such profiles are naturally generated by \kompost{} propagation, such that our results can be used in future phenomenological studies. Similarly the study of different initial conditions is left for future works. Although the pre-equilibrium spectrum will be highly sensitive to the initial composition of quarks and gluons, this is beyond the scope of this paper.

Within this work, we computed the $p_T$ differential photon spectrum, which is arguably the simplest photon observable and is not directly sensitive to the anisotropy of photon production in early times. Nevertheless, it is conceivable that other observables, such as, e.g., photon $v_2$ or HBT could be more suited to identify the unique features of pre-equilibrium production~\cite{Garcia-Montero:2019kjk}. We leave such detailed investigations for future work.

Evidently, another logical extension of our formalism is the calculation of pre-equilibrium dilepton production in QCD kinetic theory. In this context, recent studies indicate that dileptons from the pre-equilibrium phase might dominate the production in an intermediate range of the invariant mass \cite{Coquet:2021gms,Coquet:2021lca,Churchill:2020uvk,Kasmaei:2018oag}, making them an interesting candidate for further exploration of the pre-equilibrium phase of high-energy heavy-ion collisions. Preliminary results show that a similar universal scaling as for pre-equilibrium photon production can also be obtained for pre-equilibrium dilepton production, opening a new avenue for detailed phenomenological investigations. Similarly, the emergence of such universal scaling during the pre-equilibrium phase is also interesting for other phenomenological applications, concerning, e.g., heavy-flavor production, heavy-quark diffusion~\cite{Boguslavski:2023fdm,Du:2023izb} or jet energy loss during the pre-equilibrium phase, which deserve further investigations.

%

\acknowledgments
We thank J\"urgen Berges, Xiaojian Du, Charles Gale, Nicole L\"oher, Stephan Ochsenfeld, Jean-Francois Paquet and Klaus Reyers for their valuable discussions. OGM, PP and SS acknowledge support by the Deutsche Forschungsgemeinschaft (DFG, German Research Foundation) through the CRC-TR 211 ‘Strong-interaction matter under extreme conditions’-project number 315477589 – TRR 211. 
AM acknowledges support by the DFG through Emmy Noether Programme (project number 496831614)
and CRC 1225 ISOQUANT (project number 27381115).
OGM and SS acknowledge also support by the German Bundesministerium für Bildung und Forschung (BMBF) through Grant No. 05P21PBCAA. The authors acknowledge computing time provided by the Paderborn Center for Parallel Computing (PC2) and the National Energy Research Scientific Computing Center, a DOE Office of Science User Facility supported by the Office of Science of the U.S. Department of Energy under Contract No. DE-AC02-05CH11231.

\appendix
\section{Quark and gluon collision integrals in QCD kinetic theory}\label{sec:TheoryNonEqQGP}

\subsection{Elastic collision integrals}
For the collision integrals we use the notation of Arnold, Moore and Yaffe (AMY) \cite{Arnold:2002zm}. Therefore the elastic collision integral for particle $a$ with momentum $p_1$ involved in a scattering process $a,b\rightarrow c,d$ with momenta $p_1, p_2 \leftrightarrow p_3, p_4$ is given as
\begin{align}
    &C^{2 \leftrightarrow 2}_a [f_a](\pp_1) = -\frac{1}{2\nu_a} ~\frac{1}{2E_{p_1}} \\
    &\sum_{cd} \int d\Pi_{2 \leftrightarrow 2} \left| \mathcal{M}_{cd}^{ab}(\pp_1,\pp_2|\pp_3,\pp_4) \right|^2 F_{cd}^{ab}(\pp_1,\pp_2|\pp_3,\pp_4) \, , \nonumber
\end{align}
where $\mathcal{M}_{cd}^{ab}$ is the corresponding matrix element summed over spin and color, $F_{cd}^{ab}$ is the statistical factor, $\nu_g = 2(N_c^2 - 1)=16$ and $\nu_q = 2 N_c=6$. The measure is determined as
\begin{align}
    &d\Pi_{2 \leftrightarrow 2} = \frac{d^3\vec{p}_2}{(2\pi)^3} \frac{1}{2E_{p_2}} ~\frac{d^3\vec{p}_3}{(2\pi)^3} \frac{1}{2E_{p_3}} ~\frac{d^3\vec{p}_4}{(2\pi)^3} \frac{1}{2E_{p_4}} \nonumber \\
    &\times (2\pi)^4 \delta^{(4)} \left( P_1 + P_2 - P_3 - P_4 \right) \, .
\end{align}
The statistical factor $F_{cd}^{ab}$ is given as
\begin{align}
    &F_{cd}^{ab}(\pp_1,\pp_2|\pp_3,\pp_4) = f_a(\pp_1) f_b(\pp_2) (1 \pm f_c(\pp_3)) (1 \pm f_d(\pp_4)) \nonumber \\
    &- f_c(\pp_3) f_d(\pp_4) (1 \pm f_a(\pp_1)) (1 \pm f_b(\pp_2)) \, ,
\end{align}
where the $(+)$ takes into account Bose enhancement for gluons and the $(-)$ Fermi blocking for quarks.

\subsection{Inelastic collision integrals}
The inelastic collision integral for particle $a$ with momentum $P_1$ involved in a splitting process $a\rightarrow b,c$ with momenta $P_1 \leftrightarrow P_2, P_3$ and in the inverse joining process $a,b \rightarrow c$ with momenta $P_1, P_2 \leftrightarrow P_3$ has the form
\begin{align}
    &C^{1 \leftrightarrow 2}_a [f_a](\pp_1) \\
    &=- \frac{1}{2\nu_a} ~\frac{1}{2E_{p_1}} \sum_{bc} \int d\Pi_{1 \leftrightarrow 2}^{a \leftrightarrow bc} \left| \mathcal{M}_{bc}^{a}(\pp_1|\pp_2,\pp_3) \right|^2 F_{bc}^{a}(\pp_1|\pp_2,\pp_3) \nonumber \\
    &-\frac{1}{\nu_a} ~\frac{1}{2E_{p_1}} \sum_{bc} \int d\Pi_{1 \leftrightarrow 2}^{ab \leftrightarrow c} \left| \mathcal{M}_{c}^{ab}(\pp_1,\pp_2|\pp_3) \right|^2 F_{c}^{ab}(\pp_1,\pp_2|\pp_3) \nonumber \\
    &= -\frac{1}{2\nu_a} ~\frac{1}{2E_{p_1}} \sum_{bc} \int d\Pi_{1 \leftrightarrow 2}^{a \leftrightarrow bc} \left| \mathcal{M}_{bc}^{a}(\pp_1|\pp_2,\pp_3) \right|^2 F_{bc}^{a}(\pp_1|\pp_2,\pp_3) \nonumber \\
    &+\frac{1}{\nu_a} ~\frac{1}{2E_{p_1}} \sum_{bc} \int d\Pi_{1 \leftrightarrow 2}^{ab \leftrightarrow c} \left| \mathcal{M}^{c}_{ab}(\pp_3|\pp_1,\pp_2) \right|^2 F^{c}_{ab}(\pp_3|\pp_1,\pp_2) \, .
\end{align}
The measures are given by
\begin{subequations}
    \begin{align}
        &d\Pi_{1 \leftrightarrow 2}^{a \leftrightarrow bc} = \frac{d^3\vec{p}_2}{(2\pi)^3} \frac{1}{2E_{p_2}} ~\frac{d^3\vec{p}_3}{(2\pi)^3} \frac{1}{2E_{p_3}} \nonumber \\
        &\times (2\pi)^4 \delta^{(4)} \left( P_1 - P_2 - P_3 \right) \\
        &d\Pi_{1 \leftrightarrow 2}^{ab \leftrightarrow c} = \frac{d^3\vec{p}_2}{(2\pi)^3} \frac{1}{2E_{p_2}} ~\frac{d^3\vec{p}_3}{(2\pi)^3} \frac{1}{2E_{p_3}} \nonumber \\
        &\times (2\pi)^4 \delta^{(4)} \left( P_1 + P_2 - P_3 \right) \, , \nonumber
    \end{align}
\end{subequations}
while we have
\begin{subequations}
    \begin{align}
        &F_{bc}^{a}(\pp_1|\pp_2,\pp_3) = f_a(\pp_1) (1 \pm f_b(\pp_2)) (1 \pm f_c(\pp_3)) \nonumber \\
        &- f_b(\pp_2) f_c(\pp_3) (1 \pm f_a(\pp_1)) \\
        &F_{c}^{ab}(\pp_1,\pp_2|\pp_3) = f_a(\pp_1) f_b(\pp_2) (1 \pm f_c(\pp_3)) \nonumber \\
        &- f_c(\pp_3) (1 \pm f_a(\pp_1)) (1 \pm f_b(\pp_2)) \, ,
    \end{align}
\end{subequations}
for the statistical factors.

In this work we do not want to concentrate on how to solve these equations in detail but rather refer to \cite{Du:2020dvp}, where the authors do exactly this, which is the formalism we will also use. Nevertheless the dynamical evolution of distribution functions for quarks and gluons itself is important to our studies since they enter the photon production rates and is shown in \cref{fig:DistributionFunctions}.

\section{Direct photon production}
\label{sec:sourcesPheno}

In this appendix, we give details about the computation of the direct sources of photon production, previously published in Refs.~\cite{Garcia-Montero:2019kjk, Garcia-Montero:2019vju}.
\subsection{Prompt photons}

The very first photons produced in hadronic collisions are produced via hard scattering of the partons from the individual nucleons, hence the name \textit{prompt} photons. The radiation cross-section for such processes, $NN\rightarrow \gamma X$, can be computed using perturbative QCD (pQCD) \cite{Gordon:1993qc}. To compare to in-medium photon production and furthermore experimental data, we need to extend this computation to smaller $p_T$. This is done by using the same parametrization used in previous works by PHENIX and previous works on early-time photon production \cite{Garcia-Montero:2019kjk,Garcia-Montero:2019vju},
\begin{equation}
    \frac{\rmd \sigma^{pp}}{\rmd^2\pT \rmd y} = A_{pp}\left(1+\frac{p_T^2}{P_0}\right)^{-\alpha} \, .
    \label{eq:promptfit}
\end{equation}
We use this functional form to fit the pQCD data at LHC energies, which presents good agreement to data~\cite{PHENIX:2014nkk}, by fixing the renormalization scale to $\mu = p_T$. The parameters we find are $P_0=0.628$~GeV$^{2}$, $A_{pp}=0.095$~mb~GeV$^{-2}$ and $\alpha=2.375$ for ALICE at 2.76~TeV. The errorbands in \cref{fig:pTYieldALICE} are due to variation of the renormalization scale $\mu \rightarrow \{p_T/2, 2 p_T\}$.

\Cref{eq:promptfit} accounts only for the radiation from a single binary collision of two nucleons. To account for the number of prompt photons produced in a heavy-ion collision, we need to rescale the $NN\rightarrow \gamma X$ cross-section by the number of independent collisions, $N_{\rm coll}$. That is, we assume prompt photon production in nuclear collisions as a sum of a collection of $N_{\rm coll}$ incoherent binary collision events. The resultant photon spectrum is given by
\begin{equation}
\frac{\rmd N_{\rm prompt}}{\rmd^2\pT \rmd y} =\frac{N_{\rm coll}}{\sigma_{\rm inel}^{NN}}\frac{\rmd \sigma^{NN\rightarrow\gamma X}}{\rmd^2\pT \rmd y} \, ,
\end{equation}
where $\sigma^{NN}$ is the total inelastic cross-section for the collision of two nucleons. The number of collisions, $N_{\rm coll}$, was computed using the Glauber Model. For the $0-20\%$ centrality bin we find the rescaling factor to be $N_{\rm coll}/\sigma_{\rm inel}^{NN}= 19.77$~mb$^{-1}$. 

\subsection{QGP radiation and hadron gas}

To compute the thermal partonic radiation in this work, we used the well-known parametrization of the full leading order (LO) rate in thermal QGP, see Ref.~\cite{Arnold:2001ms}. This rate contains the elastic contributions, which dominate at higher momenta, but also the in-medium collinear bremsstrahlung and inelastic pair annihilation. In particular, we include the suppression of the radiation due to interference when multiple scatterings happen, i.e., the LPM effect. The transition from the QGP to hadron resonance gas photon emission rate is done at $T=160\,\text{MeV}$.

The hadronic (thermal) stage photon emission rates are taken from Ref.~\cite{Heffernan:2014mla}, where two parametrizations are given: one for the bremsstrahlung originating from $\pi\pi$ scattering dominating at lower momenta and one for the in-medium $\rho$ mesons scattering contribution. Additionally, we assume vanishing chemical potential, as we are investigating photon production at mid-rapidity in at LHC energies. The computation is performed by continuing the hydrodynamical evolution below the freeze-out temperature and computing the contribution of these emission channels down to a minimum temperature. We have computed different spectra by shifting the minimum temperature between $T=120-140$ MeV, and we have observed no change of photon spectrum in the relevant momentum range ($p_T > 0.5$~GeV).

It is known that the computation of late-time photons from the hadronic stage requires a full microscopic afterburner treatment. However, as shown in Ref.~\cite{Schafer:2021slz}, if one is interested in solely computing the emission at the level of the $p_T$ spectrum, hydrodynamical simulations have a qualitatively equal performance.

\bibliography{References}
\end{document}